\documentclass[aps,letterpaper,amsmath,prl,floatfix,twocolumn]{revtex4-1}

\usepackage{braket}
\usepackage{times}
\usepackage{color}
\usepackage{graphicx}
\usepackage{dcolumn}
\usepackage{bm}
\usepackage{subfigure}
\usepackage{hyperref}
\graphicspath{{pics/}}
\usepackage{pdfpages}

\begin{document}

\title{Ground-state phase diagram of the square lattice Hubbard model
from density matrix embedding theory}

\author{Bo-Xiao Zheng}
\author{Garnet Kin-Lic Chan}
\affiliation{Department of Chemistry, Princeton University, New Jersey 08544, United States}

\begin{abstract}
We compute the ground-state phase diagram of the Hubbard and frustrated Hubbard models 
on the square lattice with density matrix embedding theory using clusters of up to 16 sites.
We provide an error model to estimate the reliability  of the computations and complexity of
the physics at different points in the  diagram.
We find superconductivity in the ground-state as well as competition between
 inhomogeneous charge, spin, and pairing states at low doping.
The estimated errors in the study are below $T_c$ in the cuprates and on the scale of contributions in real materials that are neglected in the Hubbard model.
\end{abstract}
\maketitle


The Hubbard model~\cite{PhysRevLett.10.159,Kanamori01091963,hubbard1963electron} is one of the simplest quantum lattice models
of correlated electron materials. Its one-band realization on the square lattice
plays a central role in understanding the essential physics of
high temperature  superconductivity~\cite{Zhang1988,Dagotto1994}.
Rigorous, near exact results are available in certain limits~\cite{scalapino2007numerical}: 
at high temperatures from series expansions~\cite{PhysRevB.77.033101,PhysRevLett.97.187202,Khatami2011,Khatami2015,PhysRevB.31.4403}, 
in infinite dimensions from converged dynamical mean-field theory~\cite{georges1992hubbard,Georges1996,PhysRevB.86.125114}, and  
at weak coupling from perturbation theory~\cite{schweitzer1991weak} and renormalization group analysis~\cite{halboth2000renormalization,raghu2010superconductivity}.
Further, at half-filling, the model has no 
fermion sign problem, and unbiased determinantal quantum Monte Carlo simulations can be converged~\cite{PhysRevB.80.075116}.
Away from these limits, however, approximations are necessary.
Many numerical methods have been applied to the model at both 
finite and zero temperature,
including fixed node, 
constrained path, 
determinantal, and variational quantum Monte Carlo (QMC)~\cite{PhysRevB.58.R14685,Becca2000,PhysRevLett.72.2442,Tocchio2008,PhysRevB.55.7464,PhysRevB.78.165101,Chang2010,
  yokoyama1987variational,PhysRevB.76.180504,yamaji1998variational,PhysRevB.43.12943}, 
density matrix renormalization group (DMRG)~\cite{white2000phase,scalapino2001numerical,white2003stripes}, and dynamical cluster (DCA)~\cite{PhysRevB.58.R7475,PhysRevB.61.12739}
and (cluster) dynamical mean-field theories (CDMFT)~\cite{Lichtenstein2000,PhysRevLett.87.186401}.
These have revealed rich phenomenology in the phase diagram
 including metallic, antiferromagnetic,  d-wave (and
other kinds of) superconducting phases, a pseudogap regime, and inhomogeneous 
orders such as stripes, and charge, spin, and pair-density waves~\cite{scalapino2007numerical}.


Here, we employ density matrix embedding theory (DMET)~\cite{Knizia2012,Knizia2013}, together with
clusters of up to 16 sites and thermodynamic extrapolation, to compute a calibrated ground-state phase diagram
for the Hubbard model.
We use the term calibrated as we provide an error model 
to estimate the quality of the results, and by proxy, the complexity of the underlying physics.
The one-band (frustrated) Hubbard model on the $L \times L$ square lattice is
\begin{align}
    H=&-t\sum_{\langle ij\rangle \sigma}a_{i\sigma}^{\dagger}a_{j\sigma}-t^\prime\sum_{\langle\langle ij\rangle \rangle \sigma}a_{i\sigma}^{\dagger}a_{j\sigma}
    +U\sum_{i}n_{i\uparrow}n_{i\downarrow}
  \label{eq:Hub}
\end{align}
where $\langle \ldots\rangle$ and $\langle \langle \ldots\rangle \rangle $ denote nearest and next-nearest neighbors, respectively, $a_{i\sigma}^{(\dagger)}$ 
destroys (creates) 
a particle on site $i$ with spin $\sigma$, and $n_{i\sigma}=a_{i\sigma}^{\dagger}a_{i\sigma}$ is the number operator. 
The standard Hubbard model corresponds to $t'=0$ (we fix $t=1$).
We further study the frustrated model with  $t'=\pm 0.2$.

DMET is a cluster impurity method which is
exact for weak coupling  ($U=0$) and weak 
hybridization ($t=0$) and which becomes exact for arbitrary $U$ as the cluster
size $N_c$ increases. 
It differs from Green function impurity methods 
such as the DCA or (C)DMFT, as it is a wavefunction method, with 
a finite bath constructed to reproduce the entanglement of the cluster with
the remaining lattice sites {\it without} bath discretization error. DMET has recently been applied and 
benchmarked in a variety of settings from
lattice models~\cite{Knizia2012,PhysRevB.89.165134,PhysRevB.89.035140,PhysRevB.91.195118} to {\it ab-initio} calculations with realistic
long-range interactions~\cite{doi:10.1021/ct500512f,bulik2014electron}, and for ground-state and spectral quantities~\cite{Booth2013}. In its 
ground-state formulation, the use of wavefunctions
substantially lowers the cost
relative to Green function impurity methods, 
allowing larger clusters to become computationally affordable.


We briefly summarize the method here, with details in
the supplementary information and original references~\cite{Knizia2012,Knizia2013}. 
DMET maps the problem of solving for the bulk ground-state 
$\ket{\Psi}$ (on the $L\times L$  lattice for $L$ sufficiently large)
to solving for the ground-state of an impurity model with
$N_c$ impurity and $N_c$ bath sites.
The exact mapping is defined via the Schmidt decomposition~\cite{Peschel2012} of 
the exact $\ket{\Psi}=\sum_i \lambda_i \ket{a_i}\ket{b_i}$,
where $\{\ket{a_i}\}$ denotes impurity states, and $\{\ket{b_i}\}$, bath states.
The bulk $\ket{\Psi}$ can
be expressed exactly in the Schmidt subspace $\{\ket{a_i b_j}\}$ 
and is the ground state of the impurity Hamiltonian defined as $H_{\mathrm{imp}}=PHP$,
$P=\sum_{ij} \ket{\alpha_i \beta_j} \bra{\alpha_i \beta_j}$,
thus establishing the exact ground-state bulk to impurity mapping.
In practice, however, the exact $\ket{\Psi}$ is, of course, unknown!
DMET therefore solves an approximate impurity problem defined from a {\it model} bulk wavefunction $\ket{\Phi}$, the
ground-state of a quadratic Hamiltonian $h=h_0 +u$, where $h_0$ is
one-body part of the Hubbard Hamiltonian, and $u$ is a one-body operator 
acting in each cluster unit cell of the bulk lattice, to be determined.
Via $\ket{\Phi}$ we define the bath space, impurity Hamiltonian,
and impurity model ground-state $\ket{\Psi'}$
(which is now an approximation to  the exact bulk wavefunction $\ket{\Psi}$) and from
which energies and local observables can be measured.
Under this approximation, the bath Hilbert space spanned by $\{\ket{b_i}\}$
(of equal size to the impurity Hilbert space) becomes isomorphic to the Fock space of $N_c$ (one-particle) sites,
i.e. the bath sites.
All these quantities
are functions of the one-body operator $u$, which is determined
self-consistently by matching the one-particle density matrix of the 
impurity wavefunction $\ket{\Psi'(u)}$, and the model lattice wavefunction
$\ket{\Phi(u)}$, corresponding to the optimization
$\min_u \sum_{ij} |\braket{\Psi(u)|a^\dag_i a_j|\Psi(u)} - \braket{\Phi(u)|a^\dag_i a_j|\Phi(u)}|^2$,
where $i,j$ label impurity and bath sites. 
In this work, we used two small modifications of the original DMET procedure in Ref.~\cite{Knizia2012}.
First, we allowed $u$ to vary over particle non-conserving terms,
thus allowing $\ket{\Psi(u)}$ to spontaneously break particle number symmetry in order
to describe superconducting phases. Second, we used an additional chemical potential
on the impurity sites, to ensure that the impurity fillings for
$\ket{\Phi}$ and $\ket{\Psi'}$ exactly match. 

To obtain the ground-state phase diagram, we carried out
DMET calculations using 2$\times$2, 4$\times$2, 8$\times 2$, and 4$\times$4 impurity 
clusters, cut from a bulk square lattice with $L=72$.
We considered $t'=0, 0.2, -0.2$, and $U=2,4,6,8$, and various densities between $n=0.6-1$.
The impurity model ground-state $\ket{\Psi'}$ was determined using
 a DMRG solver with a maximum number of renormalized states $M=2000$, and which 
allowed for $U(1)$ and $SU(2)$ symmetry breaking. The
energy, local moment $m=\frac{1}{2}(n_{i\uparrow}-n_{i\downarrow})$,
double occupancy $D=\langle n_{i\uparrow} n_{i\downarrow}\rangle$, 
and local $d$-wave pairing $d_{sc}=\frac{1}{\sqrt{2}}(\langle a_{i\uparrow}a_{j\downarrow}\rangle+\langle a_{j\uparrow}a_{i\downarrow}\rangle)$ were measured 
from $\ket{\Psi'}$.

The finite cluster DMET energies and measurements
contain 3 sources of error relative to the exact thermodynamic limit. These are from
(i) errors in the DMET self-consistency, (ii) finite $M$
 in the DMRG solver (only significant for the $8\times 2$ and $4\times 4$ clusters,
 corresponding to 32 impurity plus bath sites in the impurity model), which also induces
 error in the correlation potential $u$, (iii) finite \textit{impurity cluster} size.
(The error from the use of a finite $72\times 72$ bulk lattice, 
is so small as to not affect any of the significant digits presented here). 
To estimate the thermodynamic result,
we (i) estimated DMET self-consistency quality by the convergence of expectation values in the last iterations, 
(ii) extrapolated DMRG energies and expectation values
at finite $M$ to infinite $M$, using the linear
relation with DMRG density matrix truncation error~\cite{white2007neel}, (iii) estimated the error in
$u$ due to finite $M$, by extrapolating expectation values from self-consistent $u(M)$ obtained with different solver accuracy,
(iv) extrapolated cluster size to infinite size, with
the scaling $N_c^{-1/2}$ appropriate to a non-translationally-invariant impurity. 
Each of (i) to (iv) gives an estimate of an uncertainty component (for linear extrapolations, we use the 1$\sigma$ standard deviation),
which we combined to obtain a 
single error bar on the DMET thermodynamic estimates.
Details of the error estimation and a discussion of the complete data (of
which only a fraction is presented here) are given in the supplementary information.

\begin{table}
  \caption{Ground state energy of the 2D Hubbard model. 
    All the numbers are extrapolated to the thermodynamic limit. 
    (CP-)AFQMC results are from Zhang~\cite{zhang_private}. 
    Note that the half-filling results do not involve the constrained path approximation~\cite{PhysRevLett.90.136401},
    thus is numerically exact.
    DMRG results are from White~\cite{white_private}.}
  \centering
  \begin{ruledtabular}
  \begin{tabular}{cccccc}
    U/t&Filling&DMET&AFQMC&CP-AFQMC&DMRG\\
    \hline
    2&1.0&-1.1764(3)& -1.1763(2)&-&-1.176(2)\\
    4&1.0&-0.8604(3)& -0.8603(2)&-&-0.862(2)\\
    6&1.0&-0.6561(5)& -0.6568(3)&-&-0.658(1)\\
    8&1.0&-0.5234(10)& -0.5247(2)&-&-0.5248(2)\\
    12&1.0&-0.3686(10)&-0.3693(2)&-&-0.3696(3)\\
    \hline
    4&0.8&-1.108(2)&-&-1.110(3)&-1.1040(14)\\
    4&0.6&-1.1846(5)&-&-1.185(1)&-\\
    4&0.3&-0.8800(3)&-&-0.879(1)&-\\
  \end{tabular}
  \end{ruledtabular}
  \label{tab:energy}
\end{table}

\begin{figure}[htpb]
  \includegraphics[width=\columnwidth]{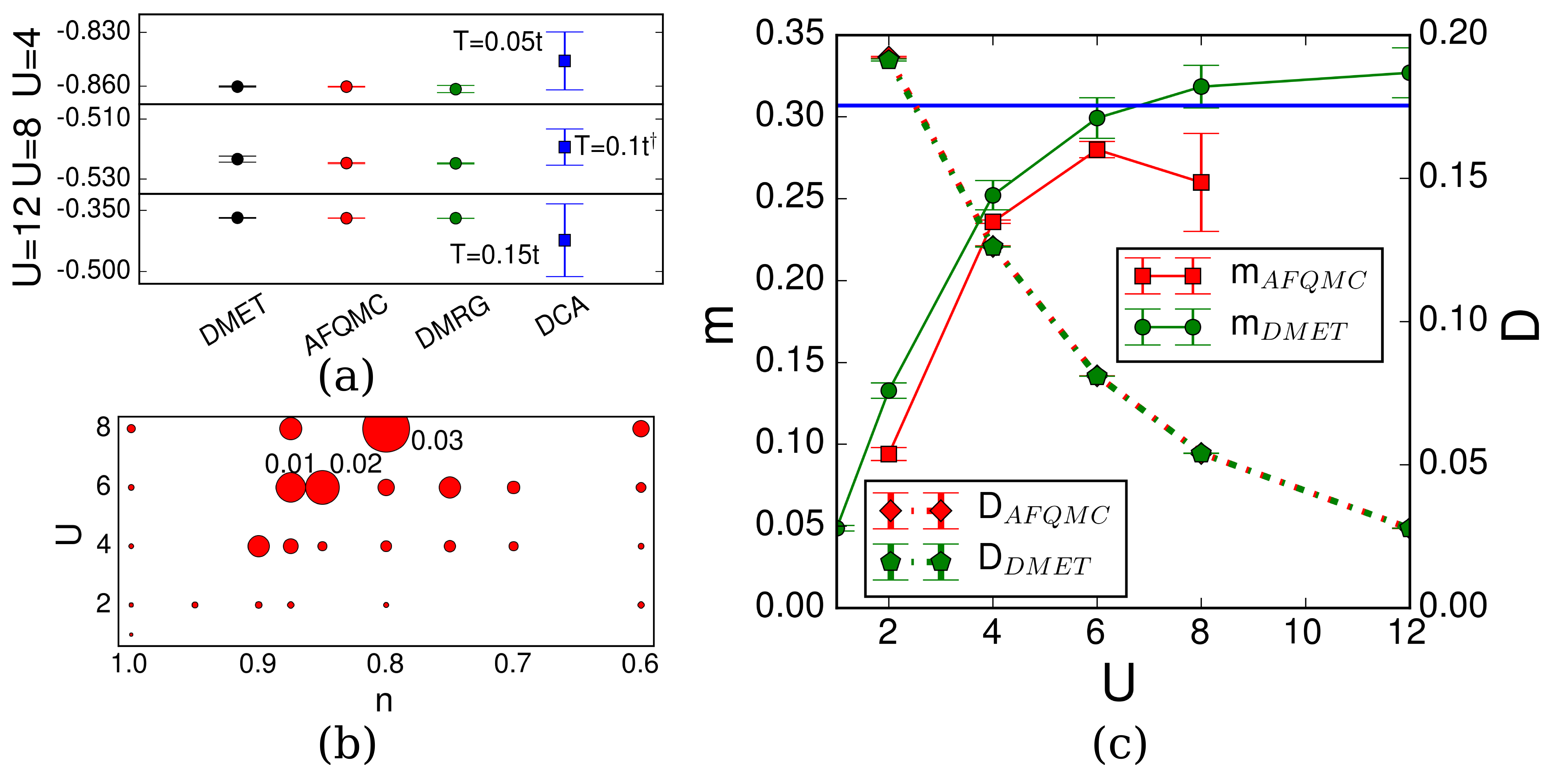}
  \caption{(color online) Benchmark and uncertainties for $t'=0$ Hubbard model. (a) Energy at half-filling. 
    Ground state estimates from DMET, AFQMC~\cite{zhang_private} and DMRG~\cite{white_private},
    compared to a recent DCA study~\cite{PhysRevB.88.155108}. The temperatures are the lowest published values
    in the DCA study. (DCA data at U=8 is from 50-site impurity cluster calculations, not extrapolated
  to thermodynamic limit.)
  (b) Energy uncertainties across the phase space. The areas of the circles are proportional
  to the estimated uncertainties.
  (c) Staggered magnetization ($m$) and double occupancy($D$) 
  at half-filling.
  The solid blue line is the spin-$\frac{1}{2}$ Heisenberg limit $m=0.3070(3)$~\cite{PhysRevB.56.11678}.}
  \label{fig:benchmark}
\end{figure}

We first verify the accuracy of our thermodynamic estimates and error bars by 
comparing to benchmark data available at half-filling.
In Table \ref{tab:energy} and Fig.~\ref{fig:benchmark},
we compare the DMET ground-state energy, double occupancies, and staggered magnetization
with exact estimates at half-filling, as obtained from ground-state (auxiliary field) 
determinantal QMC (AFQMC)
calculations on finite square lattices extrapolated to infinite size~\cite{zhang_private}, and DMRG on long open 
cylinders, extrapolated to infinite width and length~\cite{white_private}.
For comparison, we also show recent DCA energies computed 
at the lowest published temperatures, $T=0.05-0.15$t~\cite{PhysRevB.88.155108}.

The data shows the high accuracy of the DMET
energies at half-filling. The error bars from DMET, AFQMC, and DMRG are all
consistent.
Except for $U=8$ where the error is slightly larger, DMET provides the same number of significant digits as the  ``exact'' AFQMC number
with an accuracy better than 0.001$t$.
As a point of reference, the uncertainties in the
ground-state methods are significantly smaller than the finite
temperature contributions to the low-temperature DCA calculations(Fig.~\ref{fig:benchmark}(a)).

Figure~\ref{fig:benchmark}(c) further gives the half-filling staggered magnetization
and double occupancies computed with DMET, as compared
with AFQMC. The DMET double occupancies are obtained with similar error bars to the ``exact'' AFQMC
estimates. The DMET staggered magnetization, a non-local quantity, exhibits larger errors
at the smallest $U=2$ (a cluster size effect) but for $U>4$ appears similarly,
or in fact more accurate than the AFQMC result.
At the largest value $U=12$, we find $m=0.327(15)$,
slightly above the exact Heisenberg value~\cite{PhysRevB.56.11678}.


The half-filling benchmarks lend confidence to 
the DMET thermodynamic estimates of the energy and
observables, and their associated error bars. We therefore use the same 
error model to estimate the accuracy of the DMET energies and 
expectation values away from half-filling, in the absence of benchmark data.
Although exact thermodynamic limit results are not available
away from half-filling, we can verify our error model at low density using
constrained path (CP) AFQMC, a sign-free
QMC with a bias that disappears at low density and small $U$~\cite{PhysRevB.55.7464,PhysRevB.78.165101}.
For $U=4$ and $n\le0.6$, a parameter
regime where CP-AFQMC is very accurate, the DMET and CP-AFQMC energies agree to
0.001$t$ (Table~\ref{tab:energy}).
Fig.~\ref{fig:benchmark}(b) shows the energy uncertainties across the phase diagram for $t'=0$. (The
same figure for $t'=\pm 0.2$ is given in the supplementary information, which, in general, displays smaller error than $t'=0$).
As expected, the accuracy away from half-filling is significantly lower 
than at half-filling, with the largest errors found in the underdoped region of $n$=0.8-0.9.
The main source of error is from cluster size extrapolation, especially
in the underdoped region.
Large errors can be viewed as reflecting underlying physics,
as they coincide (see below) with
phase boundaries and/or the onset of 
competing inhomogeneous orders, both of which are sensitive to
cluster shape, and thus lead to errors in extrapolation.

\begin{figure}[htpb]
  \centering
  \includegraphics[width=\columnwidth]{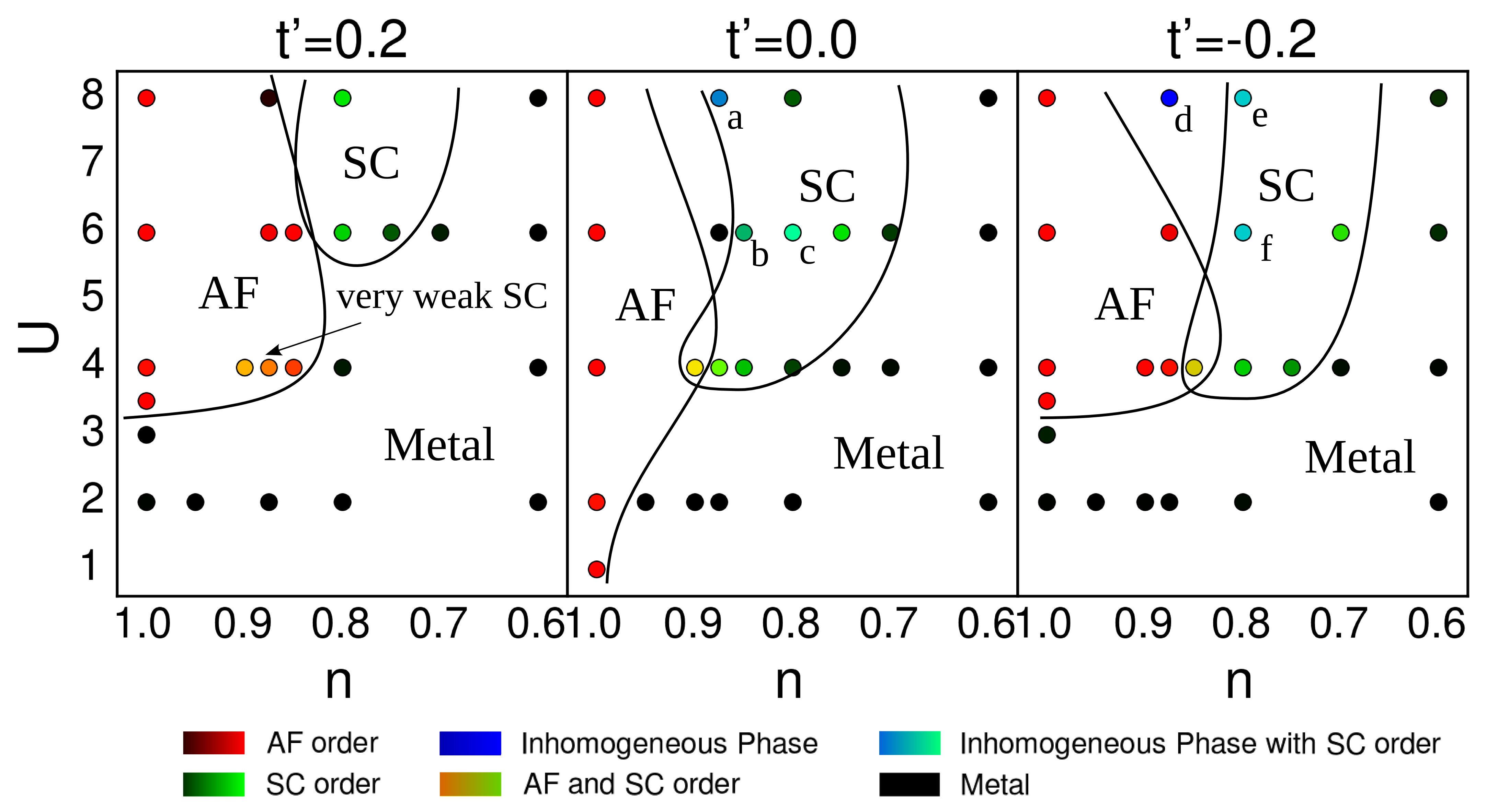}
  \caption{(color online) Phase diagrams of the standard and frustrated Hubbard model.
  Orders are represented with three primary colors: 
  red (antiferromagnetism), green (d-wave superconductivity) and blue (inhomogeneity),
  with the brightness proportional to the robustness of the order (discussed in the supplementary information).
  The points highlighted with letters: (a) local phase separation;
  (b) d-wave SC with a slight modulation in $(\pi,\pi)$ direction; (c) SC with a weak spin density wave (SDW);
  (d) a ``classic'' stripe phase; (e) stripe with pair-density wave (PDW) coexisting with SC;
  (f) CDW and spin $\pi$-phase shift; (g) intermediate points between AF and SC where both order parameters extrapolate to zero.
  \label{fig:phase}
}
\end{figure}

\begin{figure}[htpb]
  \centering
  \includegraphics[width=\columnwidth]{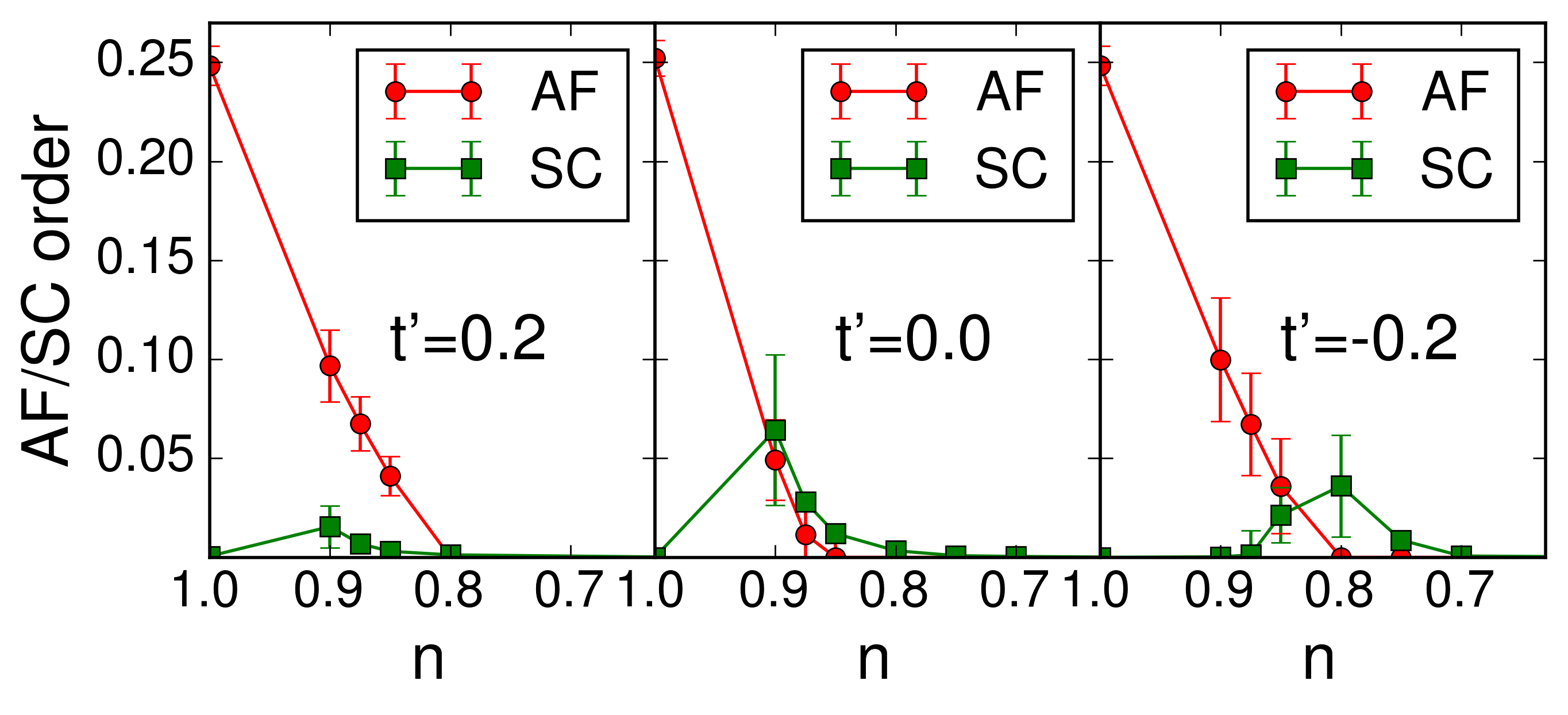}
  \caption{(color online) Antiferromagnetic (red circle) and (d-wave) superconducting (green square) order parameters at U=4.}
  \label{fig:U4order}
\end{figure}

We present the DMET phase diagrams in Fig.~\ref{fig:phase}.
We observe (i) an AF phase at half-filling, (ii) a metallic phase at large dopings and at 
small $U$, enhanced by frustration, (iii) a region of d-wave SC order at intermediate dopings and
sufficiently large $U$, (iv) a region of coexisting AF and SC order, (v) 
a region with various inhomogeneous charge, spin, and superconducting orders, (vi) points in between
the AF and SC phase where the AF and SC orders extrapolate to zero.
(The metallic phase is predicted, to be unstable at very weak coupling and large dopings from
weak coupling expansions~\cite{Metzner1989,raghu2010superconductivity}, but this
is associated with an exponentially small energy scale not probed here).
Fig. \ref{fig:U4order} shows the average AF and d-wave SC order
parameters as a function of filling for $U$=4.
We find that $t'=0.2$  stabilizes
AF versus SC, and the reverse is true for $t'=-0.2$.
For $t'=0$, the peak in SC order is around $\langle n\rangle=0.9$ and SC
extends to $\langle n\rangle\sim$0.8. The figures also
show the suppression (enhancement) of SC order with positive (negative) $t'$. 
As positive (negative) t$^\prime$ corresponds to electron-(hole-)doped cuprates,
our results are consistent with the  stronger superconductivity found in 
hole-doped materials~\cite{Pavarini2001,Huang2001,Eberlein2014}.

The presence of SC in the Hubbard model ground-state has previously been much discussed.
From the Mermin-Wagner theorem, long-range order is not allowed at finite temperatures~\cite{PhysRevLett.17.1133,0953-8984-13-27-201,Schaefer2015}, 
but at zero temperature, such long-range order can exist. In a cluster mean-field
approach embodied by cluster DMET (and similarly CDMFT)
a concern is that the observation of local order in finite clusters
does not translate into true long-range order. However, our estimates indicate that
 a homogeneous SC order parameter survives in the infinite cluster limit, within the error bars of our extrapolation.
 
At $t'=0$, we observe a banana-shaped
 SC region. At $U=6$ and $n=0.875$ (between the AF and SC phases) we find that the AF and SC order parameters are nonzero in 
 finite clusters, but extrapolate to 0 in the thermodynamic limit. However, for the analogous $U=8$,
$n=0.875$ point,  a SC state with strong inhomogeneity appears which creates large uncertainties
in the extrapolated order parameters, thus the precise location of the SC phase boundary at $U=8$ is uncertain.



\begin{figure}[thpb]
    \includegraphics[width=\columnwidth]{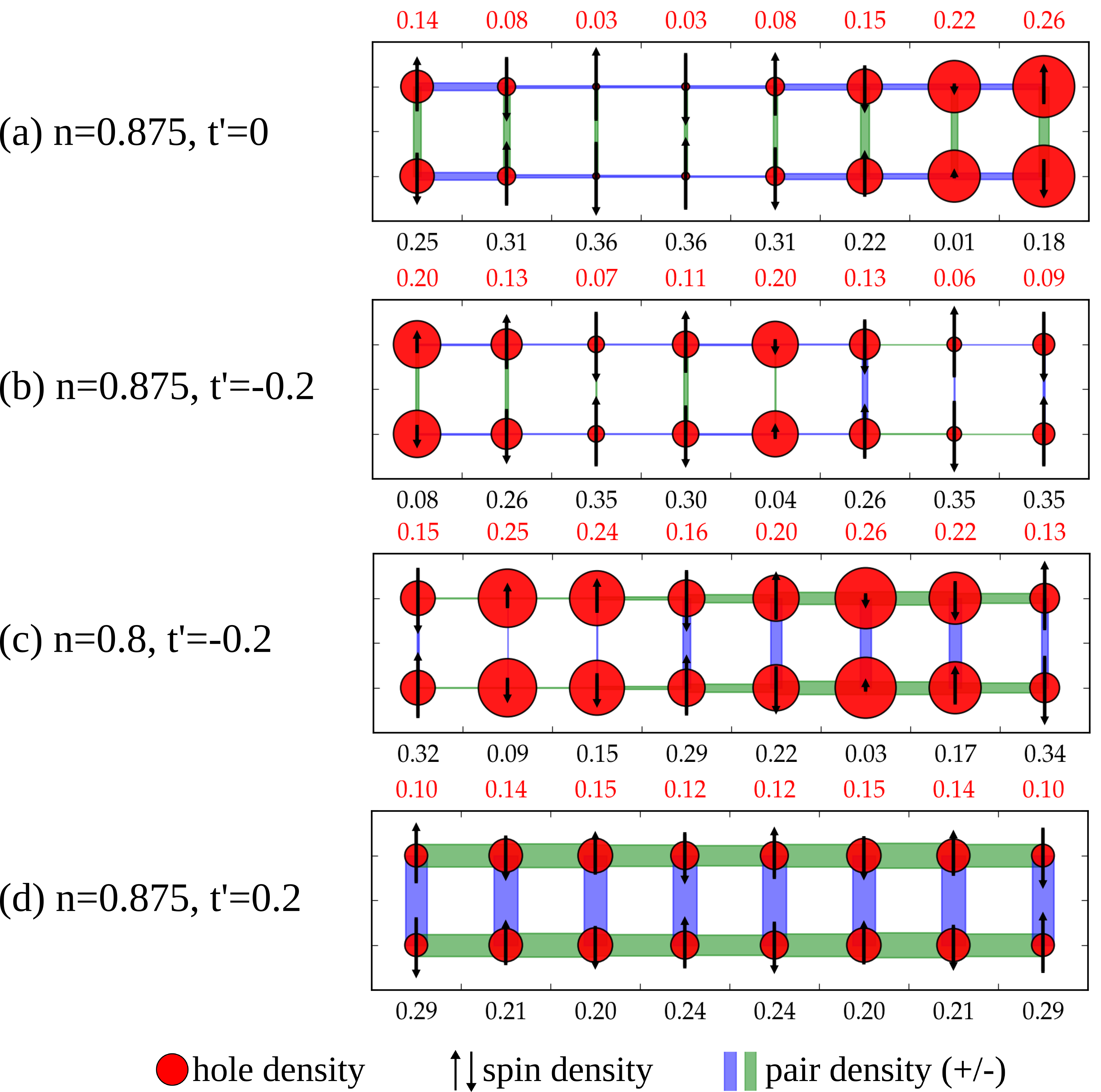}
    \caption{(color online) Local order parameters in the (frustrated) Hubbard model at selected points in the strong coupling regime ($U$=8).}
  \label{fig:stripes}
\end{figure}

We now further discuss the intermediate region between the AF and SC phases (low doping and large $U$). In this region, 
a variety of spin-density~\cite{Scalapino1986,Schulz1990,Kato1990,Chang2010,PhysRevB.89.155134,leprevost2015intertwined} 
charge-density~\cite{Poilblanc1989,Vojta1999,Melikyan2005,Chang2010}, 
pair-density wave~\cite{Chen2004a,Melikyan2005,Lee2014,berg2009charge}, and 
stripe orders~\cite{white1998density,Hellberg1999,white2000phase,white2003stripes,Hager2005,Corboz2011,leprevost2015intertwined}, have been posited in both the Hubbard model and the simpler $t$-$J$ model.
These inhomogeneous phases are proposed to be relevant in the pseudogap physics~\cite{Moshchalkov2001,Fleck2001,Chen2004a,valla2006ground,Li2006,Sedrakyan2010,Lee2014}.
Recent projected entangled pair state (PEPS) studies 
of the $t$-$J$ model suggest that different inhomogeneous 
and homogeneous states are near degenerate at low doping~\cite{Corboz2011}.
Our work indicates that the Hubbard model behaves similarly. 
For large $U$ and low doping  $n=0.875-0.8$ we find some points (marked (g) in Fig.~\ref{fig:phase}) where
the AF and SC order parameters are small or vanish, but also many other points
with various inhomogeneous orders. Some representative examples of inhomogeneous states
are shown in Fig.~\ref{fig:stripes}. These correspond to
(i) a local phase separation between a half-filled, antiferromagnetic phase and a 
superconducting ribbon (Fig.~\ref{fig:stripes}(a)),
(ii) a classic stripe phase order, with  a period 4 charge and period 8 spin density modulation (Fig.~\ref{fig:stripes}(b)),
very similar to as seen in earlier DMRG ladder studies~\cite{white2003stripes}. There is also a coexisting weak PDW
(exhibiting a sign change across the cell), consistent with earlier stripe proposals~\cite{berg2009charge}.
(iii) Inhomogeneities in the pairing order coexisting with the charge and spin orders in, eg., Fig.~\ref{fig:stripes}(c) 
where we see a PDW with
an 8 unit cell wavelength coexisting with a CDW with a 4 unit cell wavelength and a 8 unit cell SDW.
The PDW  is all positive (on the ladders) indicating coexistence with superconductivity,
similar to a recent theoretical proposal (see e.g. Ref.~\cite{Lee2014}). The inhomogeneity is mainly observed with zero or negative
$t'$, corresponding to the hole-doped cuprates. Fig.~\ref{fig:stripes}(d) shows an example with $t'=0.2$, where the inhomogeneity is much weaker than in the $t'\le0$ cases.
Although only $8\times 2$ clusters are shown above, not all
$8\times 2$ clusters are inhomogeneous, and similarly not all $4\times 4$ clusters are homogeneous.
A detailed analysis of observed
inhomogeneous phases, and the determination of the phase diagram, is presented in the supplementary information.
While the impurity clusters we use are still too small to definitively resolve the competing orders, 
they hint at the possible behaviour and energy resolution required to
determine the ground state at various points
in the phase space, and where we should focus our attention using larger clusters in future studies. 

To summarize, we have computed a ground-state phase diagram for the Hubbard and frustrated Hubbard models
on the square lattice using cluster DMET. At half-filling, the accuracy achieved by DMET appears competitive
with the best exact benchmarks, while away from half-filling our error model suggests that the calculations
remain very accurate. 
We observe standard AF and metallic phases, regions of d-wave SC pairing order, and
several kinds of inhomogeneities. At special points in the phase diagram,
the inhomogeneous and homogeneous solutions associated with $8\times 2$ and $4\times 4$ clusters
are very close in energy and definitive characterization will require higher energy resolution with larger clusters. However, for real
materials such as the cuprates, assuming $t\approx 3000$K, the energy resolution achieved here for most of the phase diagram
is already below the  superconducting transition temperature, suggesting that the near degeneracy of  these
orders will be lifted by terms beyond those in the Hubbard model, such as long range charge and hopping terms,  multi-orbital effects, and interlayer coupling. Moving beyond the 
Hubbard model to more realistic material models thus now appears of principal relevance.

\begin{acknowledgments}
  We acknowledge funding from the US Department of Energy, Office of Science, through
DE-SC0008624 and DE-SC0010530. This work was also performed as part
of the Simons Collaboration on the Many Electron Problem, sponsored by the Simons Foundation. We thank Steven White and Shiwei Zhang for
providing unpublished data, and Emanuel Gull for helpful comments. We also thank Sandeep Sharma for discussion on implementing DMRG with broken particle number symmetry.
Further discussion of the methodology and results can be found in the supplementary information.
\end{acknowledgments}

\bibliographystyle{apsrev4-1}
%


\begin{thebibliography}{78}%
\makeatletter
\providecommand \@ifxundefined [1]{%
 \@ifx{#1\undefined}
}%
\providecommand \@ifnum [1]{%
 \ifnum #1\expandafter \@firstoftwo
 \else \expandafter \@secondoftwo
 \fi
}%
\providecommand \@ifx [1]{%
 \ifx #1\expandafter \@firstoftwo
 \else \expandafter \@secondoftwo
 \fi
}%
\providecommand \natexlab [1]{#1}%
\providecommand \enquote  [1]{``#1''}%
\providecommand \bibnamefont  [1]{#1}%
\providecommand \bibfnamefont [1]{#1}%
\providecommand \citenamefont [1]{#1}%
\providecommand \href@noop [0]{\@secondoftwo}%
\providecommand \href [0]{\begingroup \@sanitize@url \@href}%
\providecommand \@href[1]{\@@startlink{#1}\@@href}%
\providecommand \@@href[1]{\endgroup#1\@@endlink}%
\providecommand \@sanitize@url [0]{\catcode `\\12\catcode `\$12\catcode
  `\&12\catcode `\#12\catcode `\^12\catcode `\_12\catcode `\%12\relax}%
\providecommand \@@startlink[1]{}%
\providecommand \@@endlink[0]{}%
\providecommand \url  [0]{\begingroup\@sanitize@url \@url }%
\providecommand \@url [1]{\endgroup\@href {#1}{\urlprefix }}%
\providecommand \urlprefix  [0]{URL }%
\providecommand \Eprint [0]{\href }%
\providecommand \doibase [0]{http://dx.doi.org/}%
\providecommand \selectlanguage [0]{\@gobble}%
\providecommand \bibinfo  [0]{\@secondoftwo}%
\providecommand \bibfield  [0]{\@secondoftwo}%
\providecommand \translation [1]{[#1]}%
\providecommand \BibitemOpen [0]{}%
\providecommand \bibitemStop [0]{}%
\providecommand \bibitemNoStop [0]{.\EOS\space}%
\providecommand \EOS [0]{\spacefactor3000\relax}%
\providecommand \BibitemShut  [1]{\csname bibitem#1\endcsname}%
\let\auto@bib@innerbib\@empty
\bibitem [{\citenamefont {Gutzwiller}(1963)}]{PhysRevLett.10.159}%
  \BibitemOpen
  \bibfield  {author} {\bibinfo {author} {\bibfnamefont {M.~C.}\ \bibnamefont
  {Gutzwiller}},\ }\href {\doibase 10.1103/PhysRevLett.10.159} {\bibfield
  {journal} {\bibinfo  {journal} {Phys. Rev. Lett.}\ }\textbf {\bibinfo
  {volume} {10}},\ \bibinfo {pages} {159} (\bibinfo {year} {1963})}\BibitemShut
  {NoStop}%
\bibitem [{\citenamefont {Kanamori}(1963)}]{Kanamori01091963}%
  \BibitemOpen
  \bibfield  {author} {\bibinfo {author} {\bibfnamefont {J.}~\bibnamefont
  {Kanamori}},\ }\href {\doibase 10.1143/PTP.30.275} {\bibfield  {journal}
  {\bibinfo  {journal} {Progress of Theoretical Physics}\ }\textbf {\bibinfo
  {volume} {30}},\ \bibinfo {pages} {275} (\bibinfo {year} {1963})}\BibitemShut
  {NoStop}%
\bibitem [{\citenamefont {Hubbard}(1963)}]{hubbard1963electron}%
  \BibitemOpen
  \bibfield  {author} {\bibinfo {author} {\bibfnamefont {J.}~\bibnamefont
  {Hubbard}},\ }in\ \href@noop {} {\emph {\bibinfo {booktitle} {Proceedings of
  the Royal Society of London A: Mathematical, Physical and Engineering
  Sciences}}},\ Vol.\ \bibinfo {volume} {276}\ (\bibinfo {organization} {The
  Royal Society},\ \bibinfo {year} {1963})\ pp.\ \bibinfo {pages}
  {238--257}\BibitemShut {NoStop}%
\bibitem [{\citenamefont {Zhang}\ and\ \citenamefont {Rice}(1988)}]{Zhang1988}%
  \BibitemOpen
  \bibfield  {author} {\bibinfo {author} {\bibfnamefont {F.~C.}\ \bibnamefont
  {Zhang}}\ and\ \bibinfo {author} {\bibfnamefont {T.~M.}\ \bibnamefont
  {Rice}},\ }\href {\doibase 10.1103/PhysRevB.37.3759} {\bibfield  {journal}
  {\bibinfo  {journal} {Phys. Rev. B}\ }\textbf {\bibinfo {volume} {37}},\
  \bibinfo {pages} {3759} (\bibinfo {year} {1988})}\BibitemShut {NoStop}%
\bibitem [{\citenamefont {Dagotto}(1994)}]{Dagotto1994}%
  \BibitemOpen
  \bibfield  {author} {\bibinfo {author} {\bibfnamefont {E.}~\bibnamefont
  {Dagotto}},\ }\href {\doibase 10.1103/RevModPhys.66.763} {\bibfield
  {journal} {\bibinfo  {journal} {Rev. Mod. Phys.}\ }\textbf {\bibinfo {volume}
  {66}},\ \bibinfo {pages} {763} (\bibinfo {year} {1994})}\BibitemShut
  {NoStop}%
\bibitem [{\citenamefont {Scalapino}(2007)}]{scalapino2007numerical}%
  \BibitemOpen
  \bibfield  {author} {\bibinfo {author} {\bibfnamefont {D.}~\bibnamefont
  {Scalapino}},\ }in\ \href@noop {} {\emph {\bibinfo {booktitle} {Handbook of
  High-Temperature Superconductivity}}}\ (\bibinfo  {publisher} {Springer},\
  \bibinfo {year} {2007})\ pp.\ \bibinfo {pages} {495--526}\BibitemShut
  {NoStop}%
\bibitem [{\citenamefont {Rubtsov}\ \emph {et~al.}(2008)\citenamefont
  {Rubtsov}, \citenamefont {Katsnelson},\ and\ \citenamefont
  {Lichtenstein}}]{PhysRevB.77.033101}%
  \BibitemOpen
  \bibfield  {author} {\bibinfo {author} {\bibfnamefont {A.~N.}\ \bibnamefont
  {Rubtsov}}, \bibinfo {author} {\bibfnamefont {M.~I.}\ \bibnamefont
  {Katsnelson}}, \ and\ \bibinfo {author} {\bibfnamefont {A.~I.}\ \bibnamefont
  {Lichtenstein}},\ }\href {\doibase 10.1103/PhysRevB.77.033101} {\bibfield
  {journal} {\bibinfo  {journal} {Phys. Rev. B}\ }\textbf {\bibinfo {volume}
  {77}},\ \bibinfo {pages} {033101} (\bibinfo {year} {2008})}\BibitemShut
  {NoStop}%
\bibitem [{\citenamefont {Rigol}\ \emph {et~al.}(2006)\citenamefont {Rigol},
  \citenamefont {Bryant},\ and\ \citenamefont {Singh}}]{PhysRevLett.97.187202}%
  \BibitemOpen
  \bibfield  {author} {\bibinfo {author} {\bibfnamefont {M.}~\bibnamefont
  {Rigol}}, \bibinfo {author} {\bibfnamefont {T.}~\bibnamefont {Bryant}}, \
  and\ \bibinfo {author} {\bibfnamefont {R.~R.~P.}\ \bibnamefont {Singh}},\
  }\href {\doibase 10.1103/PhysRevLett.97.187202} {\bibfield  {journal}
  {\bibinfo  {journal} {Phys. Rev. Lett.}\ }\textbf {\bibinfo {volume} {97}},\
  \bibinfo {pages} {187202} (\bibinfo {year} {2006})}\BibitemShut {NoStop}%
\bibitem [{\citenamefont {Khatami}\ and\ \citenamefont
  {Rigol}(2011)}]{Khatami2011}%
  \BibitemOpen
  \bibfield  {author} {\bibinfo {author} {\bibfnamefont {E.}~\bibnamefont
  {Khatami}}\ and\ \bibinfo {author} {\bibfnamefont {M.}~\bibnamefont
  {Rigol}},\ }\href {\doibase 10.1103/PhysRevA.84.053611} {\bibfield  {journal}
  {\bibinfo  {journal} {Phys. Rev. A}\ }\textbf {\bibinfo {volume} {84}},\
  \bibinfo {pages} {053611} (\bibinfo {year} {2011})}\BibitemShut {NoStop}%
\bibitem [{\citenamefont {Khatami}\ \emph {et~al.}(2015)\citenamefont
  {Khatami}, \citenamefont {Scalettar},\ and\ \citenamefont
  {Singh}}]{Khatami2015}%
  \BibitemOpen
  \bibfield  {author} {\bibinfo {author} {\bibfnamefont {E.}~\bibnamefont
  {Khatami}}, \bibinfo {author} {\bibfnamefont {R.~T.}\ \bibnamefont
  {Scalettar}}, \ and\ \bibinfo {author} {\bibfnamefont {R.~R.}\ \bibnamefont
  {Singh}},\ }\href@noop {} {\bibfield  {journal} {\bibinfo  {journal} {arXiv
  preprint arXiv:1503.06213}\ } (\bibinfo {year} {2015})}\BibitemShut {NoStop}%
\bibitem [{\citenamefont {Hirsch}(1985)}]{PhysRevB.31.4403}%
  \BibitemOpen
  \bibfield  {author} {\bibinfo {author} {\bibfnamefont {J.~E.}\ \bibnamefont
  {Hirsch}},\ }\href {\doibase 10.1103/PhysRevB.31.4403} {\bibfield  {journal}
  {\bibinfo  {journal} {Phys. Rev. B}\ }\textbf {\bibinfo {volume} {31}},\
  \bibinfo {pages} {4403} (\bibinfo {year} {1985})}\BibitemShut {NoStop}%
\bibitem [{\citenamefont {Georges}\ and\ \citenamefont
  {Kotliar}(1992)}]{georges1992hubbard}%
  \BibitemOpen
  \bibfield  {author} {\bibinfo {author} {\bibfnamefont {A.}~\bibnamefont
  {Georges}}\ and\ \bibinfo {author} {\bibfnamefont {G.}~\bibnamefont
  {Kotliar}},\ }\href@noop {} {\bibfield  {journal} {\bibinfo  {journal}
  {Physical Review B}\ }\textbf {\bibinfo {volume} {45}},\ \bibinfo {pages}
  {6479} (\bibinfo {year} {1992})}\BibitemShut {NoStop}%
\bibitem [{\citenamefont {Georges}\ \emph {et~al.}(1996)\citenamefont
  {Georges}, \citenamefont {Kotliar}, \citenamefont {Krauth},\ and\
  \citenamefont {Rozenberg}}]{Georges1996}%
  \BibitemOpen
  \bibfield  {author} {\bibinfo {author} {\bibfnamefont {A.}~\bibnamefont
  {Georges}}, \bibinfo {author} {\bibfnamefont {G.}~\bibnamefont {Kotliar}},
  \bibinfo {author} {\bibfnamefont {W.}~\bibnamefont {Krauth}}, \ and\ \bibinfo
  {author} {\bibfnamefont {M.~J.}\ \bibnamefont {Rozenberg}},\ }\href {\doibase
  10.1103/RevModPhys.68.13} {\bibfield  {journal} {\bibinfo  {journal} {Rev.
  Mod. Phys.}\ }\textbf {\bibinfo {volume} {68}},\ \bibinfo {pages} {13}
  (\bibinfo {year} {1996})}\BibitemShut {NoStop}%
\bibitem [{\citenamefont {Rohringer}\ \emph {et~al.}(2012)\citenamefont
  {Rohringer}, \citenamefont {Valli},\ and\ \citenamefont
  {Toschi}}]{PhysRevB.86.125114}%
  \BibitemOpen
  \bibfield  {author} {\bibinfo {author} {\bibfnamefont {G.}~\bibnamefont
  {Rohringer}}, \bibinfo {author} {\bibfnamefont {A.}~\bibnamefont {Valli}}, \
  and\ \bibinfo {author} {\bibfnamefont {A.}~\bibnamefont {Toschi}},\ }\href
  {\doibase 10.1103/PhysRevB.86.125114} {\bibfield  {journal} {\bibinfo
  {journal} {Phys. Rev. B}\ }\textbf {\bibinfo {volume} {86}},\ \bibinfo
  {pages} {125114} (\bibinfo {year} {2012})}\BibitemShut {NoStop}%
\bibitem [{\citenamefont {Schweitzer}\ and\ \citenamefont
  {Czycholl}(1991)}]{schweitzer1991weak}%
  \BibitemOpen
  \bibfield  {author} {\bibinfo {author} {\bibfnamefont {H.}~\bibnamefont
  {Schweitzer}}\ and\ \bibinfo {author} {\bibfnamefont {G.}~\bibnamefont
  {Czycholl}},\ }\href@noop {} {\bibfield  {journal} {\bibinfo  {journal}
  {Zeitschrift f{\"u}r Physik B Condensed Matter}\ }\textbf {\bibinfo {volume}
  {83}},\ \bibinfo {pages} {93} (\bibinfo {year} {1991})}\BibitemShut {NoStop}%
\bibitem [{\citenamefont {Halboth}\ and\ \citenamefont
  {Metzner}(2000)}]{halboth2000renormalization}%
  \BibitemOpen
  \bibfield  {author} {\bibinfo {author} {\bibfnamefont {C.~J.}\ \bibnamefont
  {Halboth}}\ and\ \bibinfo {author} {\bibfnamefont {W.}~\bibnamefont
  {Metzner}},\ }\href@noop {} {\bibfield  {journal} {\bibinfo  {journal}
  {Physical Review B}\ }\textbf {\bibinfo {volume} {61}},\ \bibinfo {pages}
  {7364} (\bibinfo {year} {2000})}\BibitemShut {NoStop}%
\bibitem [{\citenamefont {Raghu}\ \emph {et~al.}(2010)\citenamefont {Raghu},
  \citenamefont {Kivelson},\ and\ \citenamefont
  {Scalapino}}]{raghu2010superconductivity}%
  \BibitemOpen
  \bibfield  {author} {\bibinfo {author} {\bibfnamefont {S.}~\bibnamefont
  {Raghu}}, \bibinfo {author} {\bibfnamefont {S.}~\bibnamefont {Kivelson}}, \
  and\ \bibinfo {author} {\bibfnamefont {D.}~\bibnamefont {Scalapino}},\
  }\href@noop {} {\bibfield  {journal} {\bibinfo  {journal} {Physical Review
  B}\ }\textbf {\bibinfo {volume} {81}},\ \bibinfo {pages} {224505} (\bibinfo
  {year} {2010})}\BibitemShut {NoStop}%
\bibitem [{\citenamefont {Varney}\ \emph {et~al.}(2009)\citenamefont {Varney},
  \citenamefont {Lee}, \citenamefont {Bai}, \citenamefont {Chiesa},
  \citenamefont {Jarrell},\ and\ \citenamefont
  {Scalettar}}]{PhysRevB.80.075116}%
  \BibitemOpen
  \bibfield  {author} {\bibinfo {author} {\bibfnamefont {C.~N.}\ \bibnamefont
  {Varney}}, \bibinfo {author} {\bibfnamefont {C.-R.}\ \bibnamefont {Lee}},
  \bibinfo {author} {\bibfnamefont {Z.~J.}\ \bibnamefont {Bai}}, \bibinfo
  {author} {\bibfnamefont {S.}~\bibnamefont {Chiesa}}, \bibinfo {author}
  {\bibfnamefont {M.}~\bibnamefont {Jarrell}}, \ and\ \bibinfo {author}
  {\bibfnamefont {R.~T.}\ \bibnamefont {Scalettar}},\ }\href {\doibase
  10.1103/PhysRevB.80.075116} {\bibfield  {journal} {\bibinfo  {journal} {Phys.
  Rev. B}\ }\textbf {\bibinfo {volume} {80}},\ \bibinfo {pages} {075116}
  (\bibinfo {year} {2009})}\BibitemShut {NoStop}%
\bibitem [{\citenamefont {Cosentini}\ \emph {et~al.}(1998)\citenamefont
  {Cosentini}, \citenamefont {Capone}, \citenamefont {Guidoni},\ and\
  \citenamefont {Bachelet}}]{PhysRevB.58.R14685}%
  \BibitemOpen
  \bibfield  {author} {\bibinfo {author} {\bibfnamefont {A.~C.}\ \bibnamefont
  {Cosentini}}, \bibinfo {author} {\bibfnamefont {M.}~\bibnamefont {Capone}},
  \bibinfo {author} {\bibfnamefont {L.}~\bibnamefont {Guidoni}}, \ and\
  \bibinfo {author} {\bibfnamefont {G.~B.}\ \bibnamefont {Bachelet}},\ }\href
  {\doibase 10.1103/PhysRevB.58.R14685} {\bibfield  {journal} {\bibinfo
  {journal} {Phys. Rev. B}\ }\textbf {\bibinfo {volume} {58}},\ \bibinfo
  {pages} {R14685} (\bibinfo {year} {1998})}\BibitemShut {NoStop}%
\bibitem [{\citenamefont {Becca}\ \emph {et~al.}(2000)\citenamefont {Becca},
  \citenamefont {Capone},\ and\ \citenamefont {Sorella}}]{Becca2000}%
  \BibitemOpen
  \bibfield  {author} {\bibinfo {author} {\bibfnamefont {F.}~\bibnamefont
  {Becca}}, \bibinfo {author} {\bibfnamefont {M.}~\bibnamefont {Capone}}, \
  and\ \bibinfo {author} {\bibfnamefont {S.}~\bibnamefont {Sorella}},\ }\href
  {\doibase 10.1103/PhysRevB.62.12700} {\bibfield  {journal} {\bibinfo
  {journal} {Phys. Rev. B}\ }\textbf {\bibinfo {volume} {62}},\ \bibinfo
  {pages} {12700} (\bibinfo {year} {2000})}\BibitemShut {NoStop}%
\bibitem [{\citenamefont {van Bemmel}\ \emph {et~al.}(1994)\citenamefont {van
  Bemmel}, \citenamefont {ten Haaf}, \citenamefont {van Saarloos},
  \citenamefont {van Leeuwen},\ and\ \citenamefont {An}}]{PhysRevLett.72.2442}%
  \BibitemOpen
  \bibfield  {author} {\bibinfo {author} {\bibfnamefont {H.~J.~M.}\
  \bibnamefont {van Bemmel}}, \bibinfo {author} {\bibfnamefont {D.~F.~B.}\
  \bibnamefont {ten Haaf}}, \bibinfo {author} {\bibfnamefont {W.}~\bibnamefont
  {van Saarloos}}, \bibinfo {author} {\bibfnamefont {J.~M.~J.}\ \bibnamefont
  {van Leeuwen}}, \ and\ \bibinfo {author} {\bibfnamefont {G.}~\bibnamefont
  {An}},\ }\href {\doibase 10.1103/PhysRevLett.72.2442} {\bibfield  {journal}
  {\bibinfo  {journal} {Phys. Rev. Lett.}\ }\textbf {\bibinfo {volume} {72}},\
  \bibinfo {pages} {2442} (\bibinfo {year} {1994})}\BibitemShut {NoStop}%
\bibitem [{\citenamefont {Tocchio}\ \emph {et~al.}(2008)\citenamefont
  {Tocchio}, \citenamefont {Becca}, \citenamefont {Parola},\ and\ \citenamefont
  {Sorella}}]{Tocchio2008}%
  \BibitemOpen
  \bibfield  {author} {\bibinfo {author} {\bibfnamefont {L.~F.}\ \bibnamefont
  {Tocchio}}, \bibinfo {author} {\bibfnamefont {F.}~\bibnamefont {Becca}},
  \bibinfo {author} {\bibfnamefont {A.}~\bibnamefont {Parola}}, \ and\ \bibinfo
  {author} {\bibfnamefont {S.}~\bibnamefont {Sorella}},\ }\href {\doibase
  10.1103/PhysRevB.78.041101} {\bibfield  {journal} {\bibinfo  {journal} {Phys.
  Rev. B}\ }\textbf {\bibinfo {volume} {78}},\ \bibinfo {pages} {041101}
  (\bibinfo {year} {2008})}\BibitemShut {NoStop}%
\bibitem [{\citenamefont {Zhang}\ \emph {et~al.}(1997)\citenamefont {Zhang},
  \citenamefont {Carlson},\ and\ \citenamefont
  {Gubernatis}}]{PhysRevB.55.7464}%
  \BibitemOpen
  \bibfield  {author} {\bibinfo {author} {\bibfnamefont {S.}~\bibnamefont
  {Zhang}}, \bibinfo {author} {\bibfnamefont {J.}~\bibnamefont {Carlson}}, \
  and\ \bibinfo {author} {\bibfnamefont {J.~E.}\ \bibnamefont {Gubernatis}},\
  }\href {\doibase 10.1103/PhysRevB.55.7464} {\bibfield  {journal} {\bibinfo
  {journal} {Phys. Rev. B}\ }\textbf {\bibinfo {volume} {55}},\ \bibinfo
  {pages} {7464} (\bibinfo {year} {1997})}\BibitemShut {NoStop}%
\bibitem [{\citenamefont {Chang}\ and\ \citenamefont
  {Zhang}(2008)}]{PhysRevB.78.165101}%
  \BibitemOpen
  \bibfield  {author} {\bibinfo {author} {\bibfnamefont {C.-C.}\ \bibnamefont
  {Chang}}\ and\ \bibinfo {author} {\bibfnamefont {S.}~\bibnamefont {Zhang}},\
  }\href {\doibase 10.1103/PhysRevB.78.165101} {\bibfield  {journal} {\bibinfo
  {journal} {Phys. Rev. B}\ }\textbf {\bibinfo {volume} {78}},\ \bibinfo
  {pages} {165101} (\bibinfo {year} {2008})}\BibitemShut {NoStop}%
\bibitem [{\citenamefont {Chang}\ and\ \citenamefont
  {Zhang}(2010)}]{Chang2010}%
  \BibitemOpen
  \bibfield  {author} {\bibinfo {author} {\bibfnamefont {C.-C.}\ \bibnamefont
  {Chang}}\ and\ \bibinfo {author} {\bibfnamefont {S.}~\bibnamefont {Zhang}},\
  }\href {\doibase 10.1103/PhysRevLett.104.116402} {\bibfield  {journal}
  {\bibinfo  {journal} {Phys. Rev. Lett.}\ }\textbf {\bibinfo {volume} {104}},\
  \bibinfo {pages} {116402} (\bibinfo {year} {2010})}\BibitemShut {NoStop}%
\bibitem [{\citenamefont {Yokoyama}\ and\ \citenamefont
  {Shiba}(1987)}]{yokoyama1987variational}%
  \BibitemOpen
  \bibfield  {author} {\bibinfo {author} {\bibfnamefont {H.}~\bibnamefont
  {Yokoyama}}\ and\ \bibinfo {author} {\bibfnamefont {H.}~\bibnamefont
  {Shiba}},\ }\href@noop {} {\bibfield  {journal} {\bibinfo  {journal} {Journal
  of the Physical Society of Japan}\ }\textbf {\bibinfo {volume} {56}},\
  \bibinfo {pages} {1490} (\bibinfo {year} {1987})}\BibitemShut {NoStop}%
\bibitem [{\citenamefont {Eichenberger}\ and\ \citenamefont
  {Baeriswyl}(2007)}]{PhysRevB.76.180504}%
  \BibitemOpen
  \bibfield  {author} {\bibinfo {author} {\bibfnamefont {D.}~\bibnamefont
  {Eichenberger}}\ and\ \bibinfo {author} {\bibfnamefont {D.}~\bibnamefont
  {Baeriswyl}},\ }\href {\doibase 10.1103/PhysRevB.76.180504} {\bibfield
  {journal} {\bibinfo  {journal} {Phys. Rev. B}\ }\textbf {\bibinfo {volume}
  {76}},\ \bibinfo {pages} {180504} (\bibinfo {year} {2007})}\BibitemShut
  {NoStop}%
\bibitem [{\citenamefont {Yamaji}\ \emph {et~al.}(1998)\citenamefont {Yamaji},
  \citenamefont {Yanagisawa}, \citenamefont {Nakanishi},\ and\ \citenamefont
  {Koike}}]{yamaji1998variational}%
  \BibitemOpen
  \bibfield  {author} {\bibinfo {author} {\bibfnamefont {K.}~\bibnamefont
  {Yamaji}}, \bibinfo {author} {\bibfnamefont {T.}~\bibnamefont {Yanagisawa}},
  \bibinfo {author} {\bibfnamefont {T.}~\bibnamefont {Nakanishi}}, \ and\
  \bibinfo {author} {\bibfnamefont {S.}~\bibnamefont {Koike}},\ }\href@noop {}
  {\bibfield  {journal} {\bibinfo  {journal} {Physica C: Superconductivity}\
  }\textbf {\bibinfo {volume} {304}},\ \bibinfo {pages} {225} (\bibinfo {year}
  {1998})}\BibitemShut {NoStop}%
\bibitem [{\citenamefont {Giamarchi}\ and\ \citenamefont
  {Lhuillier}(1991)}]{PhysRevB.43.12943}%
  \BibitemOpen
  \bibfield  {author} {\bibinfo {author} {\bibfnamefont {T.}~\bibnamefont
  {Giamarchi}}\ and\ \bibinfo {author} {\bibfnamefont {C.}~\bibnamefont
  {Lhuillier}},\ }\href {\doibase 10.1103/PhysRevB.43.12943} {\bibfield
  {journal} {\bibinfo  {journal} {Phys. Rev. B}\ }\textbf {\bibinfo {volume}
  {43}},\ \bibinfo {pages} {12943} (\bibinfo {year} {1991})}\BibitemShut
  {NoStop}%
\bibitem [{\citenamefont {White}\ and\ \citenamefont
  {Scalapino}(2000)}]{white2000phase}%
  \BibitemOpen
  \bibfield  {author} {\bibinfo {author} {\bibfnamefont {S.~R.}\ \bibnamefont
  {White}}\ and\ \bibinfo {author} {\bibfnamefont {D.}~\bibnamefont
  {Scalapino}},\ }\href@noop {} {\bibfield  {journal} {\bibinfo  {journal}
  {Physical review B}\ }\textbf {\bibinfo {volume} {61}},\ \bibinfo {pages}
  {6320} (\bibinfo {year} {2000})}\BibitemShut {NoStop}%
\bibitem [{\citenamefont {Scalapino}\ and\ \citenamefont
  {White}(2001)}]{scalapino2001numerical}%
  \BibitemOpen
  \bibfield  {author} {\bibinfo {author} {\bibfnamefont {D.~J.}\ \bibnamefont
  {Scalapino}}\ and\ \bibinfo {author} {\bibfnamefont {S.~R.}\ \bibnamefont
  {White}},\ }\href@noop {} {\bibfield  {journal} {\bibinfo  {journal}
  {Foundations of Physics}\ }\textbf {\bibinfo {volume} {31}},\ \bibinfo
  {pages} {27} (\bibinfo {year} {2001})}\BibitemShut {NoStop}%
\bibitem [{\citenamefont {White}\ and\ \citenamefont
  {Scalapino}(2003)}]{white2003stripes}%
  \BibitemOpen
  \bibfield  {author} {\bibinfo {author} {\bibfnamefont {S.~R.}\ \bibnamefont
  {White}}\ and\ \bibinfo {author} {\bibfnamefont {D.}~\bibnamefont
  {Scalapino}},\ }\href@noop {} {\bibfield  {journal} {\bibinfo  {journal}
  {Physical review letters}\ }\textbf {\bibinfo {volume} {91}},\ \bibinfo
  {pages} {136403} (\bibinfo {year} {2003})}\BibitemShut {NoStop}%
\bibitem [{\citenamefont {Hettler}\ \emph {et~al.}(1998)\citenamefont
  {Hettler}, \citenamefont {Tahvildar-Zadeh}, \citenamefont {Jarrell},
  \citenamefont {Pruschke},\ and\ \citenamefont
  {Krishnamurthy}}]{PhysRevB.58.R7475}%
  \BibitemOpen
  \bibfield  {author} {\bibinfo {author} {\bibfnamefont {M.~H.}\ \bibnamefont
  {Hettler}}, \bibinfo {author} {\bibfnamefont {A.~N.}\ \bibnamefont
  {Tahvildar-Zadeh}}, \bibinfo {author} {\bibfnamefont {M.}~\bibnamefont
  {Jarrell}}, \bibinfo {author} {\bibfnamefont {T.}~\bibnamefont {Pruschke}}, \
  and\ \bibinfo {author} {\bibfnamefont {H.~R.}\ \bibnamefont
  {Krishnamurthy}},\ }\href {\doibase 10.1103/PhysRevB.58.R7475} {\bibfield
  {journal} {\bibinfo  {journal} {Phys. Rev. B}\ }\textbf {\bibinfo {volume}
  {58}},\ \bibinfo {pages} {R7475} (\bibinfo {year} {1998})}\BibitemShut
  {NoStop}%
\bibitem [{\citenamefont {Hettler}\ \emph {et~al.}(2000)\citenamefont
  {Hettler}, \citenamefont {Mukherjee}, \citenamefont {Jarrell},\ and\
  \citenamefont {Krishnamurthy}}]{PhysRevB.61.12739}%
  \BibitemOpen
  \bibfield  {author} {\bibinfo {author} {\bibfnamefont {M.~H.}\ \bibnamefont
  {Hettler}}, \bibinfo {author} {\bibfnamefont {M.}~\bibnamefont {Mukherjee}},
  \bibinfo {author} {\bibfnamefont {M.}~\bibnamefont {Jarrell}}, \ and\
  \bibinfo {author} {\bibfnamefont {H.~R.}\ \bibnamefont {Krishnamurthy}},\
  }\href {\doibase 10.1103/PhysRevB.61.12739} {\bibfield  {journal} {\bibinfo
  {journal} {Phys. Rev. B}\ }\textbf {\bibinfo {volume} {61}},\ \bibinfo
  {pages} {12739} (\bibinfo {year} {2000})}\BibitemShut {NoStop}%
\bibitem [{\citenamefont {Lichtenstein}\ and\ \citenamefont
  {Katsnelson}(2000)}]{Lichtenstein2000}%
  \BibitemOpen
  \bibfield  {author} {\bibinfo {author} {\bibfnamefont {A.~I.}\ \bibnamefont
  {Lichtenstein}}\ and\ \bibinfo {author} {\bibfnamefont {M.~I.}\ \bibnamefont
  {Katsnelson}},\ }\href {\doibase 10.1103/PhysRevB.62.R9283} {\bibfield
  {journal} {\bibinfo  {journal} {Phys. Rev. B}\ }\textbf {\bibinfo {volume}
  {62}},\ \bibinfo {pages} {R9283} (\bibinfo {year} {2000})}\BibitemShut
  {NoStop}%
\bibitem [{\citenamefont {Kotliar}\ \emph {et~al.}(2001)\citenamefont
  {Kotliar}, \citenamefont {Savrasov}, \citenamefont {P\'alsson},\ and\
  \citenamefont {Biroli}}]{PhysRevLett.87.186401}%
  \BibitemOpen
  \bibfield  {author} {\bibinfo {author} {\bibfnamefont {G.}~\bibnamefont
  {Kotliar}}, \bibinfo {author} {\bibfnamefont {S.~Y.}\ \bibnamefont
  {Savrasov}}, \bibinfo {author} {\bibfnamefont {G.}~\bibnamefont {P\'alsson}},
  \ and\ \bibinfo {author} {\bibfnamefont {G.}~\bibnamefont {Biroli}},\ }\href
  {\doibase 10.1103/PhysRevLett.87.186401} {\bibfield  {journal} {\bibinfo
  {journal} {Phys. Rev. Lett.}\ }\textbf {\bibinfo {volume} {87}},\ \bibinfo
  {pages} {186401} (\bibinfo {year} {2001})}\BibitemShut {NoStop}%
\bibitem [{\citenamefont {Knizia}\ and\ \citenamefont
  {Chan}(2012)}]{Knizia2012}%
  \BibitemOpen
  \bibfield  {author} {\bibinfo {author} {\bibfnamefont {G.}~\bibnamefont
  {Knizia}}\ and\ \bibinfo {author} {\bibfnamefont {G.~K.-L.}\ \bibnamefont
  {Chan}},\ }\href {\doibase 10.1103/PhysRevLett.109.186404} {\bibfield
  {journal} {\bibinfo  {journal} {Phys. Rev. Lett.}\ }\textbf {\bibinfo
  {volume} {109}},\ \bibinfo {pages} {186404} (\bibinfo {year}
  {2012})}\BibitemShut {NoStop}%
\bibitem [{\citenamefont {Knizia}\ and\ \citenamefont
  {Chan}(2013)}]{Knizia2013}%
  \BibitemOpen
  \bibfield  {author} {\bibinfo {author} {\bibfnamefont {G.}~\bibnamefont
  {Knizia}}\ and\ \bibinfo {author} {\bibfnamefont {G.~K.-L.}\ \bibnamefont
  {Chan}},\ }\href {\doibase 10.1021/ct301044e} {\bibfield  {journal} {\bibinfo
   {journal} {Journal of Chemical Theory and Computation}\ }\textbf {\bibinfo
  {volume} {9}},\ \bibinfo {pages} {1428} (\bibinfo {year} {2013})},\ \BibitemShut {NoStop}%
\bibitem [{\citenamefont {Chen}\ \emph {et~al.}(2014)\citenamefont {Chen},
  \citenamefont {Booth}, \citenamefont {Sharma}, \citenamefont {Knizia},\ and\
  \citenamefont {Chan}}]{PhysRevB.89.165134}%
  \BibitemOpen
  \bibfield  {author} {\bibinfo {author} {\bibfnamefont {Q.}~\bibnamefont
  {Chen}}, \bibinfo {author} {\bibfnamefont {G.~H.}\ \bibnamefont {Booth}},
  \bibinfo {author} {\bibfnamefont {S.}~\bibnamefont {Sharma}}, \bibinfo
  {author} {\bibfnamefont {G.}~\bibnamefont {Knizia}}, \ and\ \bibinfo {author}
  {\bibfnamefont {G.~K.-L.}\ \bibnamefont {Chan}},\ }\href {\doibase
  10.1103/PhysRevB.89.165134} {\bibfield  {journal} {\bibinfo  {journal} {Phys.
  Rev. B}\ }\textbf {\bibinfo {volume} {89}},\ \bibinfo {pages} {165134}
  (\bibinfo {year} {2014})}\BibitemShut {NoStop}%
\bibitem [{\citenamefont {Bulik}\ \emph
  {et~al.}(2014{\natexlab{a}})\citenamefont {Bulik}, \citenamefont {Scuseria},\
  and\ \citenamefont {Dukelsky}}]{PhysRevB.89.035140}%
  \BibitemOpen
  \bibfield  {author} {\bibinfo {author} {\bibfnamefont {I.~W.}\ \bibnamefont
  {Bulik}}, \bibinfo {author} {\bibfnamefont {G.~E.}\ \bibnamefont {Scuseria}},
  \ and\ \bibinfo {author} {\bibfnamefont {J.}~\bibnamefont {Dukelsky}},\
  }\href {\doibase 10.1103/PhysRevB.89.035140} {\bibfield  {journal} {\bibinfo
  {journal} {Phys. Rev. B}\ }\textbf {\bibinfo {volume} {89}},\ \bibinfo
  {pages} {035140} (\bibinfo {year} {2014}{\natexlab{a}})}\BibitemShut
  {NoStop}%
\bibitem [{\citenamefont {Fan}\ and\ \citenamefont
  {Jie}(2015)}]{PhysRevB.91.195118}%
  \BibitemOpen
  \bibfield  {author} {\bibinfo {author} {\bibfnamefont {Z.}~\bibnamefont
  {Fan}}\ and\ \bibinfo {author} {\bibfnamefont {Q.-l.}\ \bibnamefont {Jie}},\
  }\href {\doibase 10.1103/PhysRevB.91.195118} {\bibfield  {journal} {\bibinfo
  {journal} {Phys. Rev. B}\ }\textbf {\bibinfo {volume} {91}},\ \bibinfo
  {pages} {195118} (\bibinfo {year} {2015})}\BibitemShut {NoStop}%
\bibitem [{\citenamefont {Sun}\ and\ \citenamefont
  {Chan}(2014)}]{doi:10.1021/ct500512f}%
  \BibitemOpen
  \bibfield  {author} {\bibinfo {author} {\bibfnamefont {Q.}~\bibnamefont
  {Sun}}\ and\ \bibinfo {author} {\bibfnamefont {G.~K.-L.}\ \bibnamefont
  {Chan}},\ }\href {\doibase 10.1021/ct500512f} {\bibfield  {journal} {\bibinfo
   {journal} {Journal of Chemical Theory and Computation}\ }\textbf {\bibinfo
  {volume} {10}},\ \bibinfo {pages} {3784} (\bibinfo {year}
  {2014})}\BibitemShut {NoStop}%
\bibitem [{\citenamefont {Bulik}\ \emph {et~al.}(2014)\citenamefont {Bulik},
  \citenamefont {Chen},\ and\ \citenamefont {Scuseria}}]{bulik2014electron}%
  \BibitemOpen
  \bibfield  {author} {\bibinfo {author} {\bibfnamefont {I.~W.}\ \bibnamefont
  {Bulik}}, \bibinfo {author} {\bibfnamefont {W.}~\bibnamefont {Chen}}, \ and\
  \bibinfo {author} {\bibfnamefont {G.~E.}\ \bibnamefont {Scuseria}},\
  }\href@noop {} {\bibfield  {journal} {\bibinfo  {journal} {The Journal of
  chemical physics}\ }\textbf {\bibinfo {volume} {141}},\ \bibinfo {pages}
  {054113} (\bibinfo {year} {2014})}\BibitemShut {NoStop}%
\bibitem [{\citenamefont {Booth}\ and\ \citenamefont
  {Kin-Lic~Chan}(2013)}]{Booth2013}%
  \BibitemOpen
  \bibfield  {author} {\bibinfo {author} {\bibfnamefont {G.}~\bibnamefont
  {Booth}}\ and\ \bibinfo {author} {\bibfnamefont {G.}~\bibnamefont
  {Kin-Lic~Chan}},\ }\href@noop {} {\bibfield  {journal}
  }\Eprint
  {http://arxiv.org/abs/1309.2320} {arXiv:1309.2320 [cond-mat.str-el]}
  \BibitemShut {NoStop}%
\bibitem [{\citenamefont {Peschel}(2012)}]{Peschel2012}%
  \BibitemOpen
  \bibfield  {author} {\bibinfo {author} {\bibfnamefont {I.}~\bibnamefont
  {Peschel}},\ }\href {\doibase 10.1007/s13538-012-0074-1} {\bibfield
  {journal} {\bibinfo  {journal} {Brazilian Journal of Physics}\ }\textbf
  {\bibinfo {volume} {42}},\ \bibinfo {pages} {267} (\bibinfo {year}
  {2012})}\BibitemShut {NoStop}%
\bibitem [{\citenamefont {White}\ and\ \citenamefont
  {Chernyshev}(2007)}]{white2007neel}%
  \BibitemOpen
  \bibfield  {author} {\bibinfo {author} {\bibfnamefont {S.~R.}\ \bibnamefont
  {White}}\ and\ \bibinfo {author} {\bibfnamefont {A.}~\bibnamefont
  {Chernyshev}},\ }\href@noop {} {\bibfield  {journal} {\bibinfo  {journal}
  {Physical review letters}\ }\textbf {\bibinfo {volume} {99}},\ \bibinfo
  {pages} {127004} (\bibinfo {year} {2007})}\BibitemShut {NoStop}%
\bibitem [{\citenamefont {Qin}\ and\ \citenamefont
  {Zhang}(2015)}]{zhang_private}%
  \BibitemOpen
  \bibfield  {author} {\bibinfo {author} {\bibfnamefont {M.}~\bibnamefont
  {Qin}}\ and\ \bibinfo {author} {\bibfnamefont {S.}~\bibnamefont {Zhang}},\
  }\href@noop {} { {\bibinfo {title} {private communication},}\ }
  (\bibinfo {year} {2015})\BibitemShut {NoStop}%
\bibitem [{\citenamefont {Zhang}\ and\ \citenamefont
  {Krakauer}(2003)}]{PhysRevLett.90.136401}%
  \BibitemOpen
  \bibfield  {author} {\bibinfo {author} {\bibfnamefont {S.}~\bibnamefont
  {Zhang}}\ and\ \bibinfo {author} {\bibfnamefont {H.}~\bibnamefont
  {Krakauer}},\ }\href {\doibase 10.1103/PhysRevLett.90.136401} {\bibfield
  {journal} {\bibinfo  {journal} {Phys. Rev. Lett.}\ }\textbf {\bibinfo
  {volume} {90}},\ \bibinfo {pages} {136401} (\bibinfo {year}
  {2003})}\BibitemShut {NoStop}%
\bibitem [{\citenamefont {White}(2014)}]{white_private}%
  \BibitemOpen
  \bibfield  {author} {\bibinfo {author} {\bibfnamefont {S.}~\bibnamefont
  {White}},\ }\href@noop {} { {\bibinfo {title} {private
  communication},}\ } (\bibinfo {year} {2015})\BibitemShut {NoStop}%
\bibitem [{\citenamefont {LeBlanc}\ and\ \citenamefont
  {Gull}(2013)}]{PhysRevB.88.155108}%
  \BibitemOpen
  \bibfield  {author} {\bibinfo {author} {\bibfnamefont {J.}~\bibnamefont
  {LeBlanc}}\ and\ \bibinfo {author} {\bibfnamefont {E.}~\bibnamefont {Gull}},\
  }\href {\doibase 10.1103/PhysRevB.88.155108} {\bibfield  {journal} {\bibinfo
  {journal} {Phys. Rev. B}\ }\textbf {\bibinfo {volume} {88}},\ \bibinfo
  {pages} {155108} (\bibinfo {year} {2013})}\BibitemShut {NoStop}%
\bibitem [{\citenamefont {Sandvik}(1997)}]{PhysRevB.56.11678}%
  \BibitemOpen
  \bibfield  {author} {\bibinfo {author} {\bibfnamefont {A.~W.}\ \bibnamefont
  {Sandvik}},\ }\href {\doibase 10.1103/PhysRevB.56.11678} {\bibfield
  {journal} {\bibinfo  {journal} {Phys. Rev. B}\ }\textbf {\bibinfo {volume}
  {56}},\ \bibinfo {pages} {11678} (\bibinfo {year} {1997})}\BibitemShut
  {NoStop}%
\bibitem [{\citenamefont {Metzner}\ and\ \citenamefont
  {Vollhardt}(1989)}]{Metzner1989}%
  \BibitemOpen
  \bibfield  {author} {\bibinfo {author} {\bibfnamefont {W.}~\bibnamefont
  {Metzner}}\ and\ \bibinfo {author} {\bibfnamefont {D.}~\bibnamefont
  {Vollhardt}},\ }\href {\doibase 10.1103/PhysRevB.39.4462} {\bibfield
  {journal} {\bibinfo  {journal} {Phys. Rev. B}\ }\textbf {\bibinfo {volume}
  {39}},\ \bibinfo {pages} {4462} (\bibinfo {year} {1989})}\BibitemShut
  {NoStop}%
\bibitem [{\citenamefont {Pavarini}\ \emph {et~al.}(2001)\citenamefont
  {Pavarini}, \citenamefont {Dasgupta}, \citenamefont {Saha-Dasgupta},
  \citenamefont {Jepsen},\ and\ \citenamefont {Andersen}}]{Pavarini2001}%
  \BibitemOpen
  \bibfield  {author} {\bibinfo {author} {\bibfnamefont {E.}~\bibnamefont
  {Pavarini}}, \bibinfo {author} {\bibfnamefont {I.}~\bibnamefont {Dasgupta}},
  \bibinfo {author} {\bibfnamefont {T.}~\bibnamefont {Saha-Dasgupta}}, \bibinfo
  {author} {\bibfnamefont {O.}~\bibnamefont {Jepsen}}, \ and\ \bibinfo {author}
  {\bibfnamefont {O.~K.}\ \bibnamefont {Andersen}},\ }\href {\doibase
  10.1103/PhysRevLett.87.047003} {\bibfield  {journal} {\bibinfo  {journal}
  {Phys. Rev. Lett.}\ }\textbf {\bibinfo {volume} {87}},\ \bibinfo {pages}
  {047003} (\bibinfo {year} {2001})}\BibitemShut {NoStop}%
\bibitem [{\citenamefont {Huang}\ \emph {et~al.}(2001)\citenamefont {Huang},
  \citenamefont {Lin},\ and\ \citenamefont {Gubernatis}}]{Huang2001}%
  \BibitemOpen
  \bibfield  {author} {\bibinfo {author} {\bibfnamefont {Z.~B.}\ \bibnamefont
  {Huang}}, \bibinfo {author} {\bibfnamefont {H.~Q.}\ \bibnamefont {Lin}}, \
  and\ \bibinfo {author} {\bibfnamefont {J.~E.}\ \bibnamefont {Gubernatis}},\
  }\href {\doibase 10.1103/PhysRevB.64.205101} {\bibfield  {journal} {\bibinfo
  {journal} {Phys. Rev. B}\ }\textbf {\bibinfo {volume} {64}},\ \bibinfo
  {pages} {205101} (\bibinfo {year} {2001})}\BibitemShut {NoStop}%
\bibitem [{\citenamefont {Eberlein}\ and\ \citenamefont
  {Metzner}(2014)}]{Eberlein2014}%
  \BibitemOpen
  \bibfield  {author} {\bibinfo {author} {\bibfnamefont {A.}~\bibnamefont
  {Eberlein}}\ and\ \bibinfo {author} {\bibfnamefont {W.}~\bibnamefont
  {Metzner}},\ }\href {\doibase 10.1103/PhysRevB.89.035126} {\bibfield
  {journal} {\bibinfo  {journal} {Phys. Rev. B}\ }\textbf {\bibinfo {volume}
  {89}},\ \bibinfo {pages} {035126} (\bibinfo {year} {2014})}\BibitemShut
  {NoStop}%
\bibitem [{\citenamefont {Mermin}\ and\ \citenamefont
  {Wagner}(1966)}]{PhysRevLett.17.1133}%
  \BibitemOpen
  \bibfield  {author} {\bibinfo {author} {\bibfnamefont {N.~D.}\ \bibnamefont
  {Mermin}}\ and\ \bibinfo {author} {\bibfnamefont {H.}~\bibnamefont
  {Wagner}},\ }\href {\doibase 10.1103/PhysRevLett.17.1133} {\bibfield
  {journal} {\bibinfo  {journal} {Phys. Rev. Lett.}\ }\textbf {\bibinfo
  {volume} {17}},\ \bibinfo {pages} {1133} (\bibinfo {year}
  {1966})}\BibitemShut {NoStop}%
\bibitem [{\citenamefont {Gelfert}\ and\ \citenamefont
  {Nolting}(2001)}]{0953-8984-13-27-201}%
  \BibitemOpen
  \bibfield  {author} {\bibinfo {author} {\bibfnamefont {A.}~\bibnamefont
  {Gelfert}}\ and\ \bibinfo {author} {\bibfnamefont {W.}~\bibnamefont
  {Nolting}},\ }\href {http://stacks.iop.org/0953-8984/13/i=27/a=201}
  {\bibfield  {journal} {\bibinfo  {journal} {Journal of Physics: Condensed
  Matter}\ }\textbf {\bibinfo {volume} {13}},\ \bibinfo {pages} {R505}
  (\bibinfo {year} {2001})}\BibitemShut {NoStop}%
\bibitem [{\citenamefont {Sch\"afer}\ \emph {et~al.}(2015)\citenamefont
  {Sch\"afer}, \citenamefont {Geles}, \citenamefont {Rost}, \citenamefont
  {Rohringer}, \citenamefont {Arrigoni}, \citenamefont {Held}, \citenamefont
  {Bl\"umer}, \citenamefont {Aichhorn},\ and\ \citenamefont
  {Toschi}}]{Schaefer2015}%
  \BibitemOpen
  \bibfield  {author} {\bibinfo {author} {\bibfnamefont {T.}~\bibnamefont
  {Sch\"afer}}, \bibinfo {author} {\bibfnamefont {F.}~\bibnamefont {Geles}},
  \bibinfo {author} {\bibfnamefont {D.}~\bibnamefont {Rost}}, \bibinfo {author}
  {\bibfnamefont {G.}~\bibnamefont {Rohringer}}, \bibinfo {author}
  {\bibfnamefont {E.}~\bibnamefont {Arrigoni}}, \bibinfo {author}
  {\bibfnamefont {K.}~\bibnamefont {Held}}, \bibinfo {author} {\bibfnamefont
  {N.}~\bibnamefont {Bl\"umer}}, \bibinfo {author} {\bibfnamefont
  {M.}~\bibnamefont {Aichhorn}}, \ and\ \bibinfo {author} {\bibfnamefont
  {A.}~\bibnamefont {Toschi}},\ }\href {\doibase 10.1103/PhysRevB.91.125109}
  {\bibfield  {journal} {\bibinfo  {journal} {Phys. Rev. B}\ }\textbf {\bibinfo
  {volume} {91}},\ \bibinfo {pages} {125109} (\bibinfo {year}
  {2015})}\BibitemShut {NoStop}%
\bibitem [{\citenamefont {Scalapino}\ \emph {et~al.}(1986)\citenamefont
  {Scalapino}, \citenamefont {Loh},\ and\ \citenamefont
  {Hirsch}}]{Scalapino1986}%
  \BibitemOpen
  \bibfield  {author} {\bibinfo {author} {\bibfnamefont {D.~J.}\ \bibnamefont
  {Scalapino}}, \bibinfo {author} {\bibfnamefont {E.}~\bibnamefont {Loh}}, \
  and\ \bibinfo {author} {\bibfnamefont {J.~E.}\ \bibnamefont {Hirsch}},\
  }\href {\doibase 10.1103/PhysRevB.34.8190} {\bibfield  {journal} {\bibinfo
  {journal} {Phys. Rev. B}\ }\textbf {\bibinfo {volume} {34}},\ \bibinfo
  {pages} {8190} (\bibinfo {year} {1986})}\BibitemShut {NoStop}%
\bibitem [{\citenamefont {Schulz}(1990)}]{Schulz1990}%
  \BibitemOpen
  \bibfield  {author} {\bibinfo {author} {\bibfnamefont {H.~J.}\ \bibnamefont
  {Schulz}},\ }\href {\doibase 10.1103/PhysRevLett.64.1445} {\bibfield
  {journal} {\bibinfo  {journal} {Phys. Rev. Lett.}\ }\textbf {\bibinfo
  {volume} {64}},\ \bibinfo {pages} {1445} (\bibinfo {year}
  {1990})}\BibitemShut {NoStop}%
\bibitem [{\citenamefont {Kato}\ \emph {et~al.}(1990)\citenamefont {Kato},
  \citenamefont {Machida}, \citenamefont {Nakanishi},\ and\ \citenamefont
  {Fujita}}]{Kato1990}%
  \BibitemOpen
  \bibfield  {author} {\bibinfo {author} {\bibfnamefont {M.}~\bibnamefont
  {Kato}}, \bibinfo {author} {\bibfnamefont {K.}~\bibnamefont {Machida}},
  \bibinfo {author} {\bibfnamefont {H.}~\bibnamefont {Nakanishi}}, \ and\
  \bibinfo {author} {\bibfnamefont {M.}~\bibnamefont {Fujita}},\ }\href
  {\doibase 10.1143/JPSJ.59.1047} {\bibfield  {journal} {\bibinfo  {journal}
  {Journal of the Physical Society of Japan}\ }\textbf {\bibinfo {volume}
  {59}},\ \bibinfo {pages} {1047} (\bibinfo {year} {1990})}
  \BibitemShut {NoStop}%
\bibitem [{\citenamefont {Peters}\ and\ \citenamefont
  {Kawakami}(2014)}]{PhysRevB.89.155134}%
  \BibitemOpen
  \bibfield  {author} {\bibinfo {author} {\bibfnamefont {R.}~\bibnamefont
  {Peters}}\ and\ \bibinfo {author} {\bibfnamefont {N.}~\bibnamefont
  {Kawakami}},\ }\href {\doibase 10.1103/PhysRevB.89.155134} {\bibfield
  {journal} {\bibinfo  {journal} {Phys. Rev. B}\ }\textbf {\bibinfo {volume}
  {89}},\ \bibinfo {pages} {155134} (\bibinfo {year} {2014})}\BibitemShut
  {NoStop}%
\bibitem [{\citenamefont {Lepr{\'e}vost}\ \emph {et~al.}(2015)\citenamefont
  {Lepr{\'e}vost}, \citenamefont {Juillet},\ and\ \citenamefont
  {Fr{\'e}sard}}]{leprevost2015intertwined}%
  \BibitemOpen
  \bibfield  {author} {\bibinfo {author} {\bibfnamefont {A.}~\bibnamefont
  {Lepr{\'e}vost}}, \bibinfo {author} {\bibfnamefont {O.}~\bibnamefont
  {Juillet}}, \ and\ \bibinfo {author} {\bibfnamefont {R.}~\bibnamefont
  {Fr{\'e}sard}},\ }\href@noop {} {\bibfield  {journal} {\bibinfo  {journal}
  {arXiv preprint arXiv:1503.04664}\ } (\bibinfo {year} {2015})}\BibitemShut
  {NoStop}%
\bibitem [{\citenamefont {Poilblanc}\ and\ \citenamefont
  {Rice}(1989)}]{Poilblanc1989}%
  \BibitemOpen
  \bibfield  {author} {\bibinfo {author} {\bibfnamefont {D.}~\bibnamefont
  {Poilblanc}}\ and\ \bibinfo {author} {\bibfnamefont {T.~M.}\ \bibnamefont
  {Rice}},\ }\href {\doibase 10.1103/PhysRevB.39.9749} {\bibfield  {journal}
  {\bibinfo  {journal} {Phys. Rev. B}\ }\textbf {\bibinfo {volume} {39}},\
  \bibinfo {pages} {9749} (\bibinfo {year} {1989})}\BibitemShut {NoStop}%
\bibitem [{\citenamefont {Vojta}\ and\ \citenamefont
  {Sachdev}(1999)}]{Vojta1999}%
  \BibitemOpen
  \bibfield  {author} {\bibinfo {author} {\bibfnamefont {M.}~\bibnamefont
  {Vojta}}\ and\ \bibinfo {author} {\bibfnamefont {S.}~\bibnamefont
  {Sachdev}},\ }\href {\doibase 10.1103/PhysRevLett.83.3916} {\bibfield
  {journal} {\bibinfo  {journal} {Phys. Rev. Lett.}\ }\textbf {\bibinfo
  {volume} {83}},\ \bibinfo {pages} {3916} (\bibinfo {year}
  {1999})}\BibitemShut {NoStop}%
\bibitem [{\citenamefont {Melikyan}\ and\ \citenamefont {Te\ifmmode
  \check{s}\else \v{s}\fi{}anovi\ifmmode~\acute{c}\else
  \'{c}\fi{}}(2005)}]{Melikyan2005}%
  \BibitemOpen
  \bibfield  {author} {\bibinfo {author} {\bibfnamefont {A.}~\bibnamefont
  {Melikyan}}\ and\ \bibinfo {author} {\bibfnamefont {Z.}~\bibnamefont
  {Te\ifmmode \check{s}\else \v{s}\fi{}anovi\ifmmode~\acute{c}\else
  \'{c}\fi{}}},\ }\href {\doibase 10.1103/PhysRevB.71.214511} {\bibfield
  {journal} {\bibinfo  {journal} {Phys. Rev. B}\ }\textbf {\bibinfo {volume}
  {71}},\ \bibinfo {pages} {214511} (\bibinfo {year} {2005})}\BibitemShut
  {NoStop}%
\bibitem [{\citenamefont {Chen}\ \emph {et~al.}(2004)\citenamefont {Chen},
  \citenamefont {Vafek}, \citenamefont {Yazdani},\ and\ \citenamefont
  {Zhang}}]{Chen2004a}%
  \BibitemOpen
  \bibfield  {author} {\bibinfo {author} {\bibfnamefont {H.-D.}\ \bibnamefont
  {Chen}}, \bibinfo {author} {\bibfnamefont {O.}~\bibnamefont {Vafek}},
  \bibinfo {author} {\bibfnamefont {A.}~\bibnamefont {Yazdani}}, \ and\
  \bibinfo {author} {\bibfnamefont {S.-C.}\ \bibnamefont {Zhang}},\ }\href
  {\doibase 10.1103/PhysRevLett.93.187002} {\bibfield  {journal} {\bibinfo
  {journal} {Phys. Rev. Lett.}\ }\textbf {\bibinfo {volume} {93}},\ \bibinfo
  {pages} {187002} (\bibinfo {year} {2004})}\BibitemShut {NoStop}%
\bibitem [{\citenamefont {Lee}(2014)}]{Lee2014}%
  \BibitemOpen
  \bibfield  {author} {\bibinfo {author} {\bibfnamefont {P.~A.}\ \bibnamefont
  {Lee}},\ }\href {\doibase 10.1103/PhysRevX.4.031017} {\bibfield  {journal}
  {\bibinfo  {journal} {Phys. Rev. X}\ }\textbf {\bibinfo {volume} {4}},\
  \bibinfo {pages} {031017} (\bibinfo {year} {2014})}\BibitemShut {NoStop}%
\bibitem [{\citenamefont {Berg}\ \emph {et~al.}(2009)\citenamefont {Berg},
  \citenamefont {Fradkin},\ and\ \citenamefont {Kivelson}}]{berg2009charge}%
  \BibitemOpen
  \bibfield  {author} {\bibinfo {author} {\bibfnamefont {E.}~\bibnamefont
  {Berg}}, \bibinfo {author} {\bibfnamefont {E.}~\bibnamefont {Fradkin}}, \
  and\ \bibinfo {author} {\bibfnamefont {S.~A.}\ \bibnamefont {Kivelson}},\
  }\href@noop {} {\bibfield  {journal} {\bibinfo  {journal} {Nature Physics}\
  }\textbf {\bibinfo {volume} {5}},\ \bibinfo {pages} {830} (\bibinfo {year}
  {2009})}\BibitemShut {NoStop}%
\bibitem [{\citenamefont {White}\ and\ \citenamefont
  {Scalapino}(1998)}]{white1998density}%
  \BibitemOpen
  \bibfield  {author} {\bibinfo {author} {\bibfnamefont {S.~R.}\ \bibnamefont
  {White}}\ and\ \bibinfo {author} {\bibfnamefont {D.}~\bibnamefont
  {Scalapino}},\ }\href@noop {} {\bibfield  {journal} {\bibinfo  {journal}
  {Physical review letters}\ }\textbf {\bibinfo {volume} {80}},\ \bibinfo
  {pages} {1272} (\bibinfo {year} {1998})}\BibitemShut {NoStop}%
\bibitem [{\citenamefont {Hellberg}\ and\ \citenamefont
  {Manousakis}(1999)}]{Hellberg1999}%
  \BibitemOpen
  \bibfield  {author} {\bibinfo {author} {\bibfnamefont {C.~S.}\ \bibnamefont
  {Hellberg}}\ and\ \bibinfo {author} {\bibfnamefont {E.}~\bibnamefont
  {Manousakis}},\ }\href {\doibase 10.1103/PhysRevLett.83.132} {\bibfield
  {journal} {\bibinfo  {journal} {Phys. Rev. Lett.}\ }\textbf {\bibinfo
  {volume} {83}},\ \bibinfo {pages} {132} (\bibinfo {year} {1999})}\BibitemShut
  {NoStop}%
\bibitem [{\citenamefont {Hager}\ \emph {et~al.}(2005)\citenamefont {Hager},
  \citenamefont {Wellein}, \citenamefont {Jeckelmann},\ and\ \citenamefont
  {Fehske}}]{Hager2005}%
  \BibitemOpen
  \bibfield  {author} {\bibinfo {author} {\bibfnamefont {G.}~\bibnamefont
  {Hager}}, \bibinfo {author} {\bibfnamefont {G.}~\bibnamefont {Wellein}},
  \bibinfo {author} {\bibfnamefont {E.}~\bibnamefont {Jeckelmann}}, \ and\
  \bibinfo {author} {\bibfnamefont {H.}~\bibnamefont {Fehske}},\ }\href
  {\doibase 10.1103/PhysRevB.71.075108} {\bibfield  {journal} {\bibinfo
  {journal} {Phys. Rev. B}\ }\textbf {\bibinfo {volume} {71}},\ \bibinfo
  {pages} {075108} (\bibinfo {year} {2005})}\BibitemShut {NoStop}%
\bibitem [{\citenamefont {Corboz}\ \emph {et~al.}(2011)\citenamefont {Corboz},
  \citenamefont {White}, \citenamefont {Vidal},\ and\ \citenamefont
  {Troyer}}]{Corboz2011}%
  \BibitemOpen
  \bibfield  {author} {\bibinfo {author} {\bibfnamefont {P.}~\bibnamefont
  {Corboz}}, \bibinfo {author} {\bibfnamefont {S.~R.}\ \bibnamefont {White}},
  \bibinfo {author} {\bibfnamefont {G.}~\bibnamefont {Vidal}}, \ and\ \bibinfo
  {author} {\bibfnamefont {M.}~\bibnamefont {Troyer}},\ }\href {\doibase
  10.1103/PhysRevB.84.041108} {\bibfield  {journal} {\bibinfo  {journal} {Phys.
  Rev. B}\ }\textbf {\bibinfo {volume} {84}},\ \bibinfo {pages} {041108}
  (\bibinfo {year} {2011})}\BibitemShut {NoStop}%
\bibitem [{\citenamefont {Moshchalkov}\ \emph {et~al.}(2001)\citenamefont
  {Moshchalkov}, \citenamefont {Vanacken},\ and\ \citenamefont
  {Trappeniers}}]{Moshchalkov2001}%
  \BibitemOpen
  \bibfield  {author} {\bibinfo {author} {\bibfnamefont {V.~V.}\ \bibnamefont
  {Moshchalkov}}, \bibinfo {author} {\bibfnamefont {J.}~\bibnamefont
  {Vanacken}}, \ and\ \bibinfo {author} {\bibfnamefont {L.}~\bibnamefont
  {Trappeniers}},\ }\href {\doibase 10.1103/PhysRevB.64.214504} {\bibfield
  {journal} {\bibinfo  {journal} {Phys. Rev. B}\ }\textbf {\bibinfo {volume}
  {64}},\ \bibinfo {pages} {214504} (\bibinfo {year} {2001})}\BibitemShut
  {NoStop}%
\bibitem [{\citenamefont {Fleck}\ \emph {et~al.}(2001)\citenamefont {Fleck},
  \citenamefont {Lichtenstein},\ and\ \citenamefont {Ole\ifmmode~\acute{s}\else
  \'{s}\fi{}}}]{Fleck2001}%
  \BibitemOpen
  \bibfield  {author} {\bibinfo {author} {\bibfnamefont {M.}~\bibnamefont
  {Fleck}}, \bibinfo {author} {\bibfnamefont {A.~I.}\ \bibnamefont
  {Lichtenstein}}, \ and\ \bibinfo {author} {\bibfnamefont {A.~M.}\
  \bibnamefont {Ole\ifmmode~\acute{s}\else \'{s}\fi{}}},\ }\href {\doibase
  10.1103/PhysRevB.64.134528} {\bibfield  {journal} {\bibinfo  {journal} {Phys.
  Rev. B}\ }\textbf {\bibinfo {volume} {64}},\ \bibinfo {pages} {134528}
  (\bibinfo {year} {2001})}\BibitemShut {NoStop}%
\bibitem [{\citenamefont {Valla}\ \emph {et~al.}(2006)\citenamefont {Valla},
  \citenamefont {Fedorov}, \citenamefont {Lee}, \citenamefont {Davis},\ and\
  \citenamefont {Gu}}]{valla2006ground}%
  \BibitemOpen
  \bibfield  {author} {\bibinfo {author} {\bibfnamefont {T.}~\bibnamefont
  {Valla}}, \bibinfo {author} {\bibfnamefont {A.}~\bibnamefont {Fedorov}},
  \bibinfo {author} {\bibfnamefont {J.}~\bibnamefont {Lee}}, \bibinfo {author}
  {\bibfnamefont {J.}~\bibnamefont {Davis}}, \ and\ \bibinfo {author}
  {\bibfnamefont {G.}~\bibnamefont {Gu}},\ }\href@noop {} {\bibfield  {journal}
  {\bibinfo  {journal} {Science}\ }\textbf {\bibinfo {volume} {314}},\ \bibinfo
  {pages} {1914} (\bibinfo {year} {2006})}\BibitemShut {NoStop}%
\bibitem [{\citenamefont {Li}\ \emph {et~al.}(2006)\citenamefont {Li},
  \citenamefont {Wu},\ and\ \citenamefont {Lee}}]{Li2006}%
  \BibitemOpen
  \bibfield  {author} {\bibinfo {author} {\bibfnamefont {J.-X.}\ \bibnamefont
  {Li}}, \bibinfo {author} {\bibfnamefont {C.-Q.}\ \bibnamefont {Wu}}, \ and\
  \bibinfo {author} {\bibfnamefont {D.-H.}\ \bibnamefont {Lee}},\ }\href
  {\doibase 10.1103/PhysRevB.74.184515} {\bibfield  {journal} {\bibinfo
  {journal} {Phys. Rev. B}\ }\textbf {\bibinfo {volume} {74}},\ \bibinfo
  {pages} {184515} (\bibinfo {year} {2006})}\BibitemShut {NoStop}%
\bibitem [{\citenamefont {Sedrakyan}\ and\ \citenamefont
  {Chubukov}(2010)}]{Sedrakyan2010}%
  \BibitemOpen
  \bibfield  {author} {\bibinfo {author} {\bibfnamefont {T.~A.}\ \bibnamefont
  {Sedrakyan}}\ and\ \bibinfo {author} {\bibfnamefont {A.~V.}\ \bibnamefont
  {Chubukov}},\ }\href {\doibase 10.1103/PhysRevB.81.174536} {\bibfield
  {journal} {\bibinfo  {journal} {Phys. Rev. B}\ }\textbf {\bibinfo {volume}
  {81}},\ \bibinfo {pages} {174536} (\bibinfo {year} {2010})}\BibitemShut
  {NoStop}%
\end{thebibliography}
\begin{widetext}


\section*{Supplementary material}

# Supplementary information to "Ground-state phase diagram of the square lattice Hubbard model from density matrix embedding theory"

Bo-Xiao Zheng and Garnet Kin-Lic Chan

*Department of Chemistry, Princeton University, New Jersey 08544*

## I. SUMMARY OF DMET

Fig. 1 illustrates the computational flow of a DMET calculation. A DMET self-consistency cycle consists of (i) solving for the ground-state of the DMET lattice Hamiltonian, (ii) building the impurity Hamiltonian, (iii) solving for the impurity Hamiltonian ground-state and observables, and (iv) fitting the DMET correlation potential. As discussed in the main text, in this work we allow the DMET solutions to spontaneously break particle number and spin symmetry. We also include a chemical potential in the self-consistency. Here, we explain some general aspects of practical DMET calculations which have not been discussed in detail in the existing literature, as well as describe the technical extensions to broken particle number symmetry, and the self-consistency procedure for the additional chemical potential.

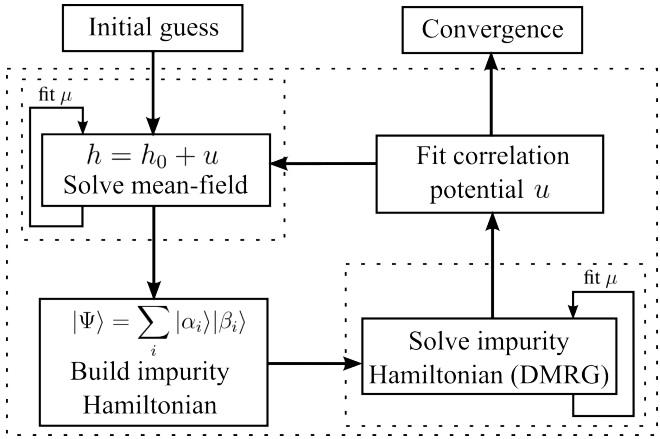

FIG. 1. Graphical representation of the DMET procedure.

### A. DMET correlation potential

A general DMET correlation potential is a quadratic operator. It is local in the sense that it does not have cross terms between different images of the impurity cluster on the lattice. In the original DMET paper[1], it took the form

$$u = \sum_C u_C = \sum_C \sum_{i,j \in C, \sigma} v_{ij} a_{i\sigma}^\dagger a_{j\sigma} \quad (1)$$

In Eq. (1), $C$ ranges over all impurity cluster supercells within the (large) lattice, $i, j$ range over sites in the same cluster $C$, and $\sigma \in \{\alpha, \beta\}$ denotes the two flavors of spin. In this form, the correlation potential has $N_c(N_c + 1)/2$ free parameters

(here and later on, we assume real potentials) where $N_c$ is the number of impurity cluster sites. For spontaneously broken particle number and spin symmetry, the correlation potential acquires additional terms,

$$u = \sum_C \sum_{i,j \in C, \sigma} v_{ij,\sigma} a_{i\sigma}^\dagger a_{j\sigma} + \Delta_{ij} a_{i\alpha}^\dagger a_{j\beta}^\dagger + h.c. \quad (2)$$

In this work, we only allow singlet pairing (strictly speaking, $S_z = 0$ pairing) but it is straightforward to extend the above to triplet pairing. The normal part $v$ has two spin components. The pairing term $\Delta$ has $N_c^2$ free parameters (it is symmetric when spin symmetry is preserved, but we allow for spin symmetry breaking). In total, the correlation potential $u$ has $N_c(2N_c + 1)$ degrees of freedom.

### B. DMET lattice Hamiltonian

The DMET lattice Hamiltonian (including a chemical potential $\mu n$) is

$$h' = h + u - \mu n = \sum_{ij\sigma} h_{ij\sigma} a_{i\sigma}^\dagger a_{j\sigma} + \Delta_{ij} a_{i\alpha}^\dagger a_{j\beta}^\dagger + c.c. \quad (3)$$

where $h$ is the hopping term in the (frustrated) Hubbard model. $h'$ can be rewritten in the form of of a spin-unrestricted Bogoliubov-de Gennes (BdG)[2,3] equation,

$$\begin{pmatrix} h'_\alpha & \Delta \\ \Delta^T & -h'_\beta \end{pmatrix} \begin{pmatrix} U_\alpha \\ V_\beta \end{pmatrix} = \begin{pmatrix} U_\alpha \\ V_\beta \end{pmatrix} \varepsilon_\alpha$$

$$\begin{pmatrix} h'_\beta & -\Delta^T \\ -\Delta & -h'_\alpha \end{pmatrix} \begin{pmatrix} U_\beta \\ V_\alpha \end{pmatrix} = \begin{pmatrix} U_\beta \\ V_\alpha \end{pmatrix} \varepsilon_\beta \quad (4)$$

These coupled equations are expressed concisely as

$$\begin{pmatrix} h'_\alpha & \Delta \\ \Delta^T & -h'_\beta \end{pmatrix} \begin{pmatrix} U_\alpha & V_\alpha \\ V_\beta & U_\beta \end{pmatrix} = \begin{pmatrix} U_\alpha & V_\alpha \\ V_\beta & U_\beta \end{pmatrix} \begin{pmatrix} \varepsilon_\alpha & \\ & -\varepsilon_\beta \end{pmatrix} \quad (5)$$

where $\varepsilon_\alpha$ and $\varepsilon_\beta$ are both positive. $h'$ is diagonalized by transforming to the Bogoliubov quasiparticles,

$$c_{i\alpha}^\dagger = u_{ji}^\alpha a_{j\alpha}^\dagger + v_{ji}^\beta a_{j\beta}$$

$$c_{i\beta}^\dagger = u_{ji}^\beta a_{j\beta}^\dagger + v_{ji}^\alpha a_{j\alpha} \quad (6)$$

Note that the number of $\{c_\alpha^\dagger\}$ and $\{c_\beta^\dagger\}$ quasiparticles will differ if $S_z \neq 0$ in the physical ground-state.

In terms of the quasiparticles, the lattice Hamiltonian in Eq. (3) is diagonalized as

$$h' = E_0 + \sum_{i\sigma} \varepsilon_{i\sigma} c_{i\sigma}^\dagger c_{i\sigma} \quad (7)$$

and the (ground state) quasiparticle vacuum $|-\rangle$, defined by $c_{i\sigma}|-\rangle = 0$, has energy $E_0$. The quasiparticle vacuum is also known as the Bardeen-Cooper-Schrieffer (BCS) ground-state[4].

### C. DMET impurity model Schmidt subspace

We now discuss how to define the impurity model Schmidt subspace corresponding to a BCS ground-state of the lattice Hamiltonian in Eq. (3). To start, we review the "product space" construction of the impurity model Schmidt subspace, starting from the lattice Hamiltonian Slater determinant ground-state, as used in the original DMET[1,5].

The original DMET impurity model consists of a set of impurity sites augmented by a set of bath modes. In Ref.[5], the bath modes are defined through the *projected overlap matrix* of the Slater determinant. We compute the projected overlap matrix from the Slater determinant coefficient matrix,

$$C_0 = \begin{pmatrix} M \\ N \end{pmatrix}_{L \times n} \qquad (8)$$

where the rows denote physical sites, and columns are occupied modes (orbitals). The upper part $M$ has $N_c$ rows. The projected overlap matrix is

$$S = M^T M \qquad (9)$$

From the singular value decomposition (SVD) of $M$ as $M = L\Sigma R^T$ (where we use the "full" form of the SVD, $L$ is $N_c \times N_c$, $\Sigma$ is $N_c \times n$, and $R$ is $n \times n$) then $S = R(\Sigma^T \Sigma)R^T$, i.e $R$ is the eigenvector matrix of the projected overlap matrix. $R$ defines a unitary transformation of the occupied modes in $C_0$, giving a new coefficient matrix $C = C_0 R$, where

$$C = \begin{pmatrix} L\Sigma \\ NR \end{pmatrix} = \begin{pmatrix} A & 0 \\ B & D \end{pmatrix} \qquad (10)$$

and the second equality follows because $\Sigma$ is a rectangular matrix of the form $(\text{diag}(\sigma), \mathbf{0}, \mathbf{0}, \ldots \mathbf{0})$, where the first $N_c$ columns constitute a diagonal matrix, and the remaining $n - N_c$ columns are zero columns. The first $N_c$ columns of $C$, $\begin{pmatrix} A \\ B \end{pmatrix}$ define the *embedding modes*, which have non-zero weight on the impurity sites. The matrix $B$ defines the *bath modes*, which may be orthonormalized using the $QR$ decomposition, $B = QR$. The remaining columns in $C$ define the *core modes*, which have no weight on the impurity. The Schmidt subspace is then $\mathcal{F}(a_i^\dagger) \otimes \mathcal{F}(b_i^\dagger) \otimes |e_1 \ldots e_{n-N_c}\rangle$, where $\{a_i^\dagger\}$ create the impurity site basis, $\{b_i^\dagger\}$ create the bath modes (from the columns of $Q$), and $|e_1 \ldots e_{n-N_c}\rangle$ is the core state, defined by the columns of $D$. The coefficients defining $\{a_i^\dagger\}$, $\{b_i^\dagger\}$ can be gathered in the columns of a matrix $C_1$,

$$C_1 = \begin{pmatrix} I_{N_c} \\ Q \end{pmatrix} \qquad (11)$$

where $I_{N_c}$ is an $N_c \times N_c$ identity matrix.

$A$ and $B$ can also be obtained directly from the one-particle density matrix. The rotation between $C$ and $C_0$ leaves the one-body density matrix invariant, thus

$$\begin{aligned} \rho = \langle a_i^\dagger a_j \rangle &= C_0 C_0^T = CC^T \\ &= \begin{pmatrix} AA^T & AB^T \\ BA^T & BB^T + DD^T \end{pmatrix} \equiv \begin{pmatrix} \rho_{\text{imp}} & \rho_c^T \\ \rho_c & \rho_{\text{env}} \end{pmatrix} \end{aligned} \qquad (12)$$

Defining the eigendecomposition $\rho_{\text{imp}} = U\Lambda U^T$, we find

$$A = U\Lambda^{\frac{1}{2}} \text{ and } B = \rho_c(A^T)^{-1} \qquad (13)$$

The above defines the impurity model Schmidt subspace as a tensor product of the impurity site space, and a bath space, thus we refer to it as a "product-space" embedding construction. However, for the BCS state, it is easier to use a slightly different, but equivalent construction. We explain this first for the Slater determinant. Here we build a $L \times 2N_c$ matrix $C_2$, whose columns span the same vector space as $C_1$ in Eq. (11), but which does not have the block structure. We start with the 'hole' one-particle density matrix

$$\rho_h = \langle a_i a_j^\dagger \rangle = I - \rho^T = I - \rho \qquad (14)$$

We can replace $\rho$ with $\rho_h$ in Eqs. (12) and (13) and compute an analogous set of coefficients $A'$ and $B'$. Taking $A$, $B$, and $A'$, $B'$ gives $C_2$,

$$C_2 = \begin{pmatrix} A & A' \\ B & B' \end{pmatrix}_{L \times 2N_c} \qquad (15)$$

The $2N_c$ columns of $C_2$ span *exactly the same space* as $C_1$ (proved in the Appendix). Thus, we can equivalently define the Schmidt subspace from the columns of $C_2$ as we can from $C_1$. Transforming to the quasiparticle vacuum of the Slater determinant, $|-\rangle$, the columns of $C_2$ define a set of $2N_c$ quasiparticle creation operators

$$c_{i\sigma}^\dagger = \sum_{j \in imp} A_{ji} a_{j\sigma'} + \sum_{j \in env} B_{ji} a_{j\sigma'} \qquad (16)$$

$$c_{\bar{i}\sigma}^\dagger = \sum_{j \in imp} A'_{ji} a_{j\sigma}^\dagger + \sum_{j \in env} B'_{ji} a_{j\sigma}^\dagger \qquad (17)$$

that yields the Schmidt subspace as $\mathcal{F}(\{c_{i(\bar{i})\sigma}^\dagger\}) \otimes |-\rangle$. As the impurity model Schmidt subspace here does not (transparently) separate between the impurity sites and environment sites, but rather involves a set of modes which are a linear transformation of both the occupied and virtual modes in the Slater determinant, we refer to this as a "quasiparticle embedding" construction. This provides an alternative view of the DMET embedding as an active space method that uses the embedding quasiparticles defined from $C_2$ as the active space, while freezing other excitations that involve only the environment.

Extending the quasiparticle embedding construction to BCS states is straightforward. By analogy with the one-particle density matrix of a Slater determinant, we define the generalized one-body density matrix for BCS states,

$$G_\sigma = \begin{pmatrix} U_\sigma \\ V_{\bar{\sigma}} \end{pmatrix} \begin{pmatrix} U_\sigma^T & V_{\bar{\sigma}}^T \end{pmatrix} = \begin{pmatrix} 1 - \rho_\sigma & \kappa_\sigma \\ \kappa_\sigma^T & \rho_{\bar{\sigma}} \end{pmatrix} \qquad (18)$$

where the normal one-particle density matrices $\rho_\sigma = \langle a_{i\sigma}^\dagger a_{j\sigma} \rangle = V_\sigma V_\sigma^T = 1 - U_\sigma U_\sigma^T$, and the pairing density matrix $\kappa = \langle a_{i\sigma} a_{j\bar\sigma} \rangle = \kappa_\alpha = -\kappa_\beta^T = U_\alpha V_\beta^T$. The diagonal of $G$ is formed by the hole and particle density matrices, and the off-diagonals by the pairing matrix. When the BCS state is a Slater determinant, $\kappa = 0$.

We reorganize the generalized density matrix $G$ into impurity and environment blocks, placing the impurity (environment) submatrices of $\rho$ and $\kappa$ together.

$$G = \begin{pmatrix} G_{\text{imp}} & G_c^T \\ G_c & G_{\text{env}} \end{pmatrix} \qquad (19)$$

Then, similar to the treatment in Eq. (13), we rewrite the impurity part of the density matrix $G_{\text{imp}} = \bar A \bar A^T$, and define a quasiparticle $C_2$ matrix

$$C_2 = \begin{pmatrix} \bar U_{\sigma,\text{imp}} \\ \bar V_{\bar\sigma,\text{imp}} \\ \bar U_{\sigma,\text{env}} \\ \bar V_{\bar\sigma,\text{env}} \end{pmatrix} = \begin{pmatrix} G_{\text{imp}} \\ G_c \end{pmatrix} (\bar A^T)^{-1} = \begin{pmatrix} \bar A \\ \bar B \end{pmatrix}_{2L \times 2N_c} \qquad (20)$$

where $\bar B = G_c(\bar A^T)^{-1}$. Eq. (20) defines a new set of quasiparticles (with associated quasiparticle creation operators $\{\bar c_{i\sigma}^\dagger\}$) in Eq. (6) through the coefficients $\bar U_\sigma, \bar V_\sigma$. These are a unitary rotation of the original $2L$ quasiparticles such that only $2N_c$ of them have non-zero overlap with the impurity. As the rotation does not mix the quasiparticle creation and annihilation operators, the vacuum of $\bar c_{i\sigma}$ is still the BCS ground state $|-\rangle$. In analogy to the embedding for Slater determinants, the Schmidt subspace is now spanned by the embedding quasiparticles, $\mathcal{F}(\{\bar c_{i\sigma}^\dagger\}) \otimes |-\rangle$.

To connect with Eq. (15), note that when the BCS state is a Slater determinant, $G_{\text{imp}}$ and $G_c$ are both block diagonal, and thus, $\bar A = \text{diag}(A'^\sigma, A^{\bar\sigma})$, $\bar B = \text{diag}(B'^\sigma, B^{\bar\sigma})$, and Eq. (20) becomes

$$\bar U_\sigma = \begin{pmatrix} A'^\sigma & 0 \\ B'^\sigma & 0 \end{pmatrix} \qquad (21)$$

$$\bar V_{\bar\sigma} = \begin{pmatrix} 0 & A^\sigma \\ 0 & B^\sigma \end{pmatrix} \qquad (22)$$

Combining both sets of spins, the quasiparticles in Eq. (22) then span exactly the same Hilbert space as the basis defined in Eq. (15). For general BCS ground states, however, $\bar A$ and $\bar B$ are not block diagonal, and the embedding quasiparticles are mixtures of particles and holes.

### D. DMET impurity Hamiltonian and DMRG solver

Once the Schmidt subspace has been defined, the DMET Hamiltonian is formally obtained by projecting an interacting lattice Hamiltonian into the subspace as $H_{\text{imp}} = PH'P$, with the many-particle projector defined as

$$P = \sum_{\vec n_{i\sigma}} |\Psi_{\vec n_{i\sigma}}\rangle \langle \Psi_{\vec n_{i\sigma}}|, \qquad (23)$$

where $\vec n_{i\sigma}$ is a vector of occupation numbers of the embedding quasiparticles, and $|\Psi_{\vec n_{i\sigma}}\rangle = \prod_{n_{i\sigma}} (c_{i\sigma}^\dagger)^{n_{i\sigma}}|-\rangle$. In earlier DMET work, two choices of lattice Hamiltonian were used in the projection: the original interacting lattice Hamiltonian $H$ (in this case the original Hubbard Hamiltonian) and a modified interacting lattice Hamiltonian $H'$, where the interaction term $U$ is only used in the impurity sites. As in earlier DMET work on lattice models, here we use the latter simpler Anderson-like lattice Hamiltonian $H'$. In $H'$, on the environment sites (outside of the impurity cluster) the Coulomb interaction $U$ is replaced by the correlation potential $u$, giving

$$H' = h + \sum_{C \neq \text{imp}} u_C + \sum_{i \in \text{imp}} U n_{i\alpha} n_{i\beta} - \mu n \qquad (24)$$

The projection defined in Eq. (23) reduces to transforming $\{a_{i\sigma}^\dagger\}$ to the embedding quasiparticles using the inverse Bogoliubov transformation,

$$a_{i\sigma}^\dagger = u_{ij}^\sigma c_{j\sigma} + v_{ij}^\sigma c_{j\bar\sigma} \qquad (25)$$

and replacing the pure environment quasiparticle operators with their expectation values with the BCS ground state $|-\rangle$.

After projection, we can write $H_{\text{imp}}$ as a sum of one- and two-particle parts, $H_{\text{imp}} = h_{\text{imp}} + V_{\text{imp}}$, where $h_{\text{imp}}$ is

$$h_{\text{imp}} = \bar h_{ij}^\sigma c_{i\sigma}^\dagger c_{j\sigma} + \bar\Delta_{ij} c_{i\alpha}^\dagger c_{j\beta}^\dagger + c.c. + E_0 \qquad (26)$$

and $c_{i\sigma}^{(\dagger)}$ here denote the embedding quasiparticles. The two-particle part $V$ contains many contributions due to the breaking of particle number symmetry in the quasiparticle formulation. These have the form

$$\begin{aligned} V_{\text{imp}} = &\frac{1}{2} \sum_{pqsr,\sigma\mu} w_{pqsr,\sigma\mu} c_{p\sigma}^\dagger c_{q\mu}^\dagger c_{s\mu} c_{r\sigma} \\ &+ \sum_{pq,\sigma} h_{pq,\sigma} c_{p\sigma}^\dagger c_{q\sigma} + E_1 \\ &+ \frac{1}{4} \sum_{pqsr} x_{pqsr} c_{p\alpha}^\dagger c_{q\alpha}^\dagger c_{s\beta}^\dagger c_{r\beta}^\dagger \\ &+ \frac{1}{2} \sum_{pqsr,\sigma} \tilde v_{pqsr,\sigma} c_{p\sigma}^\dagger c_{q\sigma}^\dagger c_{s\bar\sigma}^\dagger c_{r\sigma} \\ &+ \sum_{pq} \Delta_{pq} c_{p\alpha}^\dagger c_{q\beta}^\dagger + c.c. \end{aligned} \qquad (27)$$

$V_{\text{imp}}$ connects $N$ electron states with $N, N \pm 2, N \pm 4$ states. For brevity, we do not give the formula for the coefficients explicitly (which are obtained by simple algebra from Eq. (25)). The scalar and one-particle terms in $V_{\text{imp}}$ contain contributions from pure environment quasiparticles, and can be absorbed into $h_{\text{imp}}$.

We have adapted our quantum chemistry density matrix renormalization group (DMRG) code BLOCK[7–9] to break $U(1)$ particle number symmetry and to incorporate the Hamiltonian terms in Eq. (26) and (27). While the full wavefunction is not restricted to $U(1)$ symmetry, the particle quantum number is still used in the calculation in the sense that the renormalized states are required to carry a definite particle number.

This allows us to use the block-sparsity of the operators to tackle larger numbers of renormalized states.

The quasi-particle basis associated with $c_{i\sigma}^{(\dagger)}$ is not localized to a site, thus we use a localization and ordering procedure as used in quantum chemistry DMRG calculations to reduce long-range entanglement between the embedding quasiparticles. We find that the localization and reordering significantly reduce the DMRG truncation error, by up to a factor of 10.

### E. Expectation values

As discussed in the original papers on DMET[5], the DMET energy of $H_{imp}$ defined in Eq. (26) does not correspond to the ground-state energy of the impurity cluster. This is because the impurity Hamiltonian contains three types of energy contributions: pure impurity, impurity-bath interactions, and pure bath (environment) parts. The proper DMET energy should exclude the pure environment contributions and include only part of the impurity-bath interaction energy. Therefore, the DMET energy is evaluated as a *partial* trace of the one- and two-particle reduced density matrices of the impurity wavefunction. This partial trace can be equivalently implemented as a full trace, with appropriate scaling factors for terms in the Hamiltonian which couple the impurity and environment. For each class of term in the Hamiltonian, this scaling factor is given by the number of indices in the impurity, divided by the total number of indices. (For example, for the one-particle terms in the Hamiltonian, the contribution of the impurity-bath block to the total trace is scaled by a factor of $\frac{1}{2}$.)

An equivalent formulation for the Hubbard Hamiltonian (which contains no long-range Coulomb terms) is to evaluate the two-particle part of DMET energy as

$$E_2 = \langle \Psi | V_{imp} | \Psi \rangle = E_{DMRG} - \langle \Psi | h_{imp} | \Psi \rangle, \qquad (28)$$

where $|\Psi\rangle$ is the DMRG ground state. Since $h_{imp}$ is a quadratic operator, $E_2$ can be computed only with knowledge of the DMRG energy and the one-particle (and pairing) density matrix, avoiding explicitly evaluating $\langle \Psi | V_{imp} | \Psi \rangle$ through the two-particle density matrix.

The local spin moments and pairing are both one-particle quantities. We therefore obtain them from the one-particle and pairing density matrix $\rho = \langle c_{i\sigma}^\dagger c_{j\sigma} \rangle$, $\kappa = \langle c_{i\alpha} c_{j\beta} \rangle$ of the DMRG wavefunction $|\Psi\rangle$, transformed back to the lattice site basis $\{a_{i\sigma}^{(\dagger)}\}$ using Eq. (25). Note that $\rho$ and $\kappa$ are defined not only for quasiparticles inside the impurity Schmidt subspace, but also for core quasiparticles. (In the quasiparticle approach, although $\rho$ and $\kappa$ are themselves zero in the core, terms such as $c_i c_j^\dagger$ can appear in the expansion using Eq. (25) and result in non-zero expectation values due to the contributions of the new vacuum.) If one is interested only in impurity cluster expectation values, or for DMET lattice Hamiltonians without broken symmetry, the contribution of the core quasiparticles is strictly zero and may thus be omitted. However, for ordered (e.g. magnetic or superconducting) states, the core contribution does not vanish an therefore *cannot* be neglected. Doing so would produce for example, the strange result of vanishing

long-range correlations even in an long-range ordered DMET state.

In this study, when a single value of the order parameter is given, it is computed using the $2 \times 2$ plaquette at the center of the impurity cluster, to minimize the boundary effects. The antiferromagnetic order parameter is defined as

$$m = \frac{1}{4}(m_{0,0} + m_{1,1} - m_{0,1} - m_{1,0}) \qquad (29)$$

and the d-wave parameter as

$$d = \frac{1}{4}[d_{(0,0),(0,1)} + d_{(1,0),(1,1)} - d_{(0,0),(1,0)} - d_{(0,1),(1,1)}] \qquad (30)$$

where $m_i = \frac{1}{2}(n_{i\alpha} - n_{i\beta})$ and $d_{ij} = \frac{1}{\sqrt{2}}(\langle a_{i\alpha} a_{j\beta} \rangle + \langle a_{j\alpha} a_{i\beta} \rangle)$ as defined in the main text. Of course in some cases there are also inhomogeneous states. When the inhomogeneity is strong, we report here the full distribution of local order parameters.

### F. DMET self-consistency

The DMET embedding constructs the impurity model via the model ground-state of the DMET lattice Hamiltonian, however, this state (and the lattice Hamiltonian) are functions of the correlation potential $u$. $u$ is determined by the self-consistency procedure, which aims to minimize the difference between the embedding wavefunction and the DMET mean-field wavefunction, as measured by their (generalized) one-particle density matrix difference. In the quasiparticle embedding space, the one-particle and pairing density matrices of the mean-field wavefunction $|\Phi\rangle$ are simply zero. Conceptually, the simplest technique is to define $u$ to minimize the Frobenius norm,

$$\min_u \|G_{\Psi(u)} - G_{\Phi(u)}\|_F$$
$$= \min_u \sum_{ij} (|\rho_{\Psi,ij}^\alpha|^2 + |\rho_{\Psi,ij}^\beta|^2 + 2|\kappa_{\Psi,ij}|^2) \qquad (31)$$

However, as the derivative of the correlated wavefunction $|\Psi\rangle$ with respect to $u$ is expensive, the above is solved in a two-step procedure consisting of an inner and outer loop. In the inner loop, we carry out $\min_u \|G_\Psi - G_{\Phi(u)}\|_F$, i.e. the correlated wavefunction is held fixed, while in the outer loop, the updated $u$ leads to a new impurity model, and a new correlated wavefunction $\Psi$.

If the total particle number $n$ is allowed to fluctuate, as in a superconducting state, then one of the conjugate pairs (chemical potential) $\mu$ or (particle density) $\langle n \rangle$ must be fixed. We usually want to express the observables as a function of doping, or occupation, thus we fix $\langle n \rangle$ and determine the appropriate $\mu$. Since the diagonal elements of the correlation and chemical potential appear redundant, how can one determine the chemical potential? Formally, at the DMET mean-field level (Eq. (3)), there is a gauge freedom between $u$ and $\mu$, namely

$$\mu' = \mu + \phi, u' = u + \phi \sum_{i\sigma} a_{i\sigma}^\dagger a_{i\sigma} \qquad (32)$$

however, this gauge freedom is lost at the embedding stage (Eq. (24)), because $u$ is only added to the environment (sites outside of the impurity) while $\mu$ affects *every* site in the lattice, including the impurity. This difference allows us to use the two-step self-consistency scheme to determine $\mu$, as shown in Fig. 1. Specifically, we first fit $\mu$ at the mean-field stage, to ensure $\langle n \rangle$ is correct. Then at the embedding stage, we vary $\mu$ and $u$ simultaneously, following Eq. (32). This means that the DMET mean-field solution (and thus definition of the impurity model) stay the same, but the relative energy levels of the impurity change as compared to the bath states, which allows us to adjust the filling on the impurity.

Fitting at the embedding stage means we need to solve the correlated impurity problem more than once in a single DMET self-consistency iteration. This increases the computational cost. Our strategy is to allow only one iteration of chemical potential fitting in each DMET iteration, corresponding to at most three DMRG calculations. Because fitting $\mu$ is a one dimensional search, even with this crude approach, we can usually control the relative deviation of $\langle n \rangle$ to under $10^{-4}$.

## II. ERROR MODEL

As described in the main text, we consider 3 sources of error: (i) errors in DMET self-consistency, (ii) finite $M$ in the DMRG solver, and (iii) finite impurity cluster size. The DMET self-consistency error is estimated as $\frac{1}{2}|E^{(n-1)} - E^{(n)}|$, where $E^{(n)}$ and $E^{(n-1)}$ are the energies of the last two DMET self-consistency iterations. A typical DMET calculation oscillates between two slightly different solutions with the magnitude of the oscillations decreasing with the number of iterations. We use the range of oscillation as a representation of the self-consistency error. The error distributions across the range of calculations in this work are shown in Fig. 2, with the average values in the inset. For most points in the phase diagram, and for all cluster sizes, the self-consistency error is less than $0.0005t$. For $4 \times 4$ clusters DMET calculations are the harder to converge, giving a largest error of up to $0.002t$, and an average self-consistency error approximately twice as large as that for the other cluster shapes.

For impurity clusters larger than the $2 \times 2$ cluster (where our DMRG solver essentially does an exact diagonalization), there is error due to using finite $M$ in the DMRG impurity solver. The error due to finite $M$ has two components:

1. variational error in the DMRG calculation, which is usually assumed proportional to the density matrix truncation weight $\delta_w$,

2. the DMET correlation potential error $\delta_u$, as $\delta_u$ is a function of the impurity density matrices, and these have an error for finite $M$.

For the $4 \times 2$ and $8 \times 2$ clusters, $\delta_u$ appears negligible. For these clusters, we carry out the DMET self-consistency with lower $M$ to obtain the DMET correlation potential $u$, then do a few final DMRG calculations at large $M$ to extrapolate to the

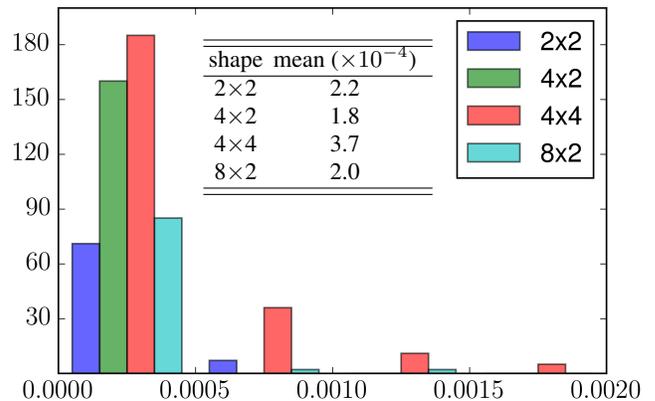

FIG. 2. Distribution and average value (inset) of the DMET self-consistency error in the energy (units of $t$) for each cluster size.

$M \to \infty$ exact solver limit. For $4 \times 4$ clusters, the $U = 2$ data is processed in this way as well. However, for other values of $U$ using the $4 \times 4$ clusters, the DMRG truncation weight is as large as $10^{-3}$ for low to intermediate doping with our accessible $M$, thus making the contribution of $\delta_u$ also significant. To compensate for this, we first carry out the DMET self-consistency with a series of different $M$ up to 1200, and linearly extrapolate the energy to the $M = \infty$ limit, $E_1$. This thus extrapolates errors from *both* source 1 and 2, assuming $\delta_u \propto \delta_w$. Another further set of DMRG calculations are then done with $M$ up to 2000, using the converged correlation potential from the DMET self-consistency with the largest $M$. This second set of results are then extrapolated again against the truncated weight to obtain an energy $E_2$, which only accounts for the error from source 1. Although the linear relation between the source 2 error and the truncation weight need not hold in general, in practice, we find that $\delta_u = \frac{1}{2}|E_1 - E_2|$ gives a crude estimate of $\delta_u$. Therefore, we report the $4 \times 4$ cluster energy as $E_{4\times4} = \frac{1}{2}(E_1 + E_2)$, with a final uncertainty of $\delta E_{4\times4}^2 = \delta_u^2 + \delta E_1^2 + \delta E_2^2$, where $\delta E_1$ is a combination of the linear regression uncertainty and the uncertainties of the original data points (from DMET self-consistency error), while $E_2$ does not have any self-consistency error. Fig. 3 illustrates the set of computations and linear extrapolations performed with each $4 \times 4$ cluster to obtain the $4 \times 4$ cluster energy and error estimate.

After obtaining the energy and observables for each cluster size, we extrapolate to the thermodynamic limit using the relation $\Delta E_{N_c} \propto N_c^{-1/2}$. Since both the $4 \times 4$ and $2 \times 8$ clusters are 16 site clusters, we must choose which one to use in the extrapolation to the thermodynamic limit. We believe that $4 \times 4$ clusters have less finite size error than the $8 \times 2$ clusters, and thus we generally use these in the extrapolation. However, at certain points in the phase diagram (e.g. at strong coupling, or negative $t'$) there is a strong tendency to inhomogeneity, and the $4 \times 4$ clusters cannot accommodate the new order parameters, resulting in a much higher energy. In such cases, namely, when (a) $4 \times 4$ and $8 \times 2$ clusters show different orders, and (b) the $8 \times 2$ cluster is lower in energy, we use the

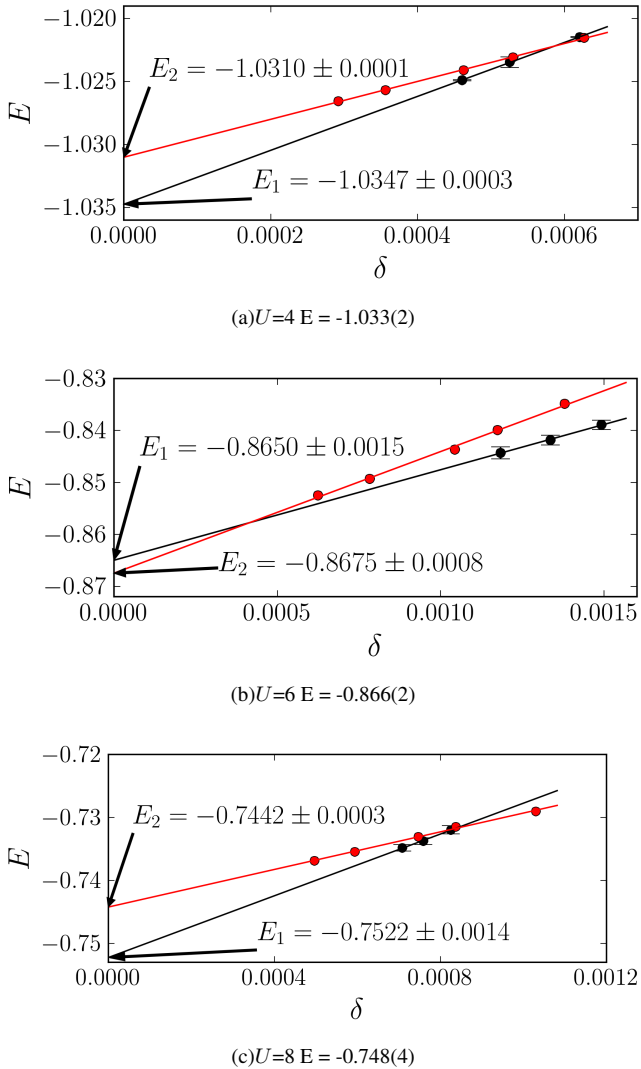

(a)$U$=4 E = -1.033(2)

(b)$U$=6 E = -0.866(2)

(c)$U$=8 E = -0.748(4)

FIG. 3. Computations involved in the estimate of the $4 \times 4$ cluster DMET energy. The black dots (with error bars for the self-consistency error) are DMET self-consistent results using different DMRG $M$. These points are extrapolated to obtain $E_1$. The red dots are DMRG results using the "best" self-consistent correlation potential, which are then extrapolated to obtain $E_2$. The final $4 \times 4$ cluster DMET energies are reported as $E = \frac{1}{2}(E_1 + E_2)$. The plots are shown for $t' = 0$, $n = 0.875$ and (a) $U$=4 (b) $U$=6 (c) $U$=8.

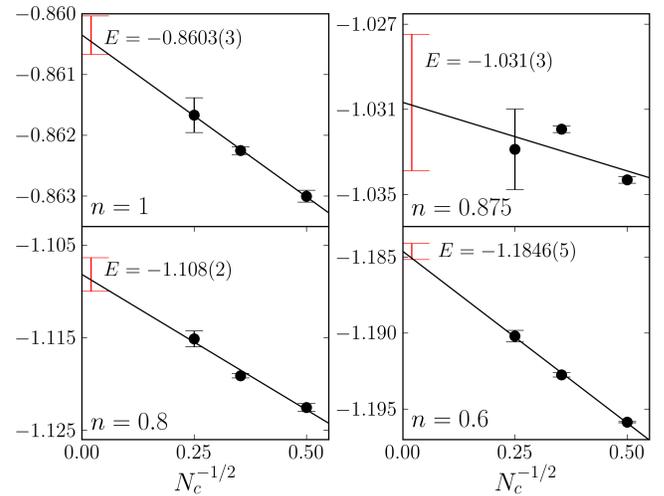

FIG. 4. Cluster size extrapolation for $U = 4, t' = 0$ at various fillings. The black dots are finite size results. The red error bars are the confidence intervals for the thermodynamic limit.

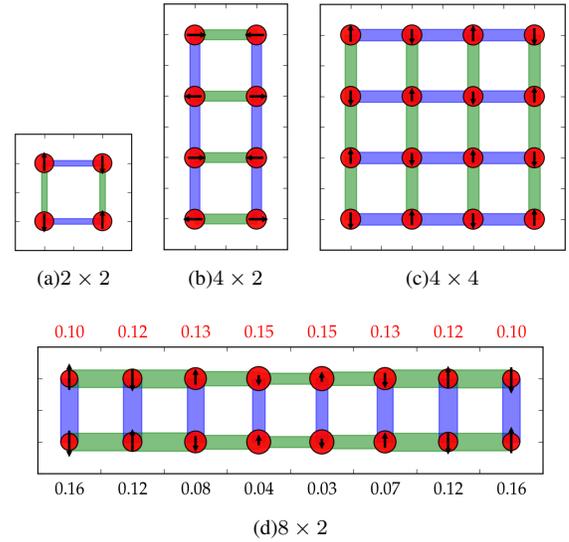

(a)$2 \times 2$    (b)$4 \times 2$    (c)$4 \times 4$

(d)$8 \times 2$

FIG. 5. Local order parameters for $U = 4, n = 0.875, t' = 0$. The legend is the same as for Fig. 5 in the main text.

$8 \times 2$ cluster energy for the extrapolation.

The cluster size extrapolation works surprisingly well given the limited number and small sizes of the clusters, although it contributes the main source of error in the final uncertainty. In Fig. 4 we show some of the extrapolation results at $U = 4$. At half-filling and in the overdoped region ($n < 0.8$), the linear relation used in the cluster size extrapolation appears quite good even for these small clusters. In the underdoped region, however, the energy is more strongly dependent on the cluster shape, often because the system has a strong tendency to establish an inhomogenous phase. In Fig. 5, we plot the local order parameters at $n = 0.875$, where the $8 \times 2$ clus-

ter calculation gives an incommensurate antiferromagnetic order. Although the $8 \times 2$ cluster energy ($-1.0288$) is slightly higher than the $4 \times 4$ cluster result ($-1.033$), its inhomogeneity suggests the existence of a low-lying inhomogeneous state that can be (relatively) stabilized by special cluster shapes. Nonetheless, even in the underdoped region, the error model appears to give a reliable estimate of the energy at the thermodynamic limit, albeit with a large uncertainty.

Fig. 6 shows the final energy errors for $t' = \pm 0.2$ across the phase diagram. The same plot for $t' = 0$ is shown in Fig. 2 in the main text. The overall uncertainty for $t' = 0.2$ is smaller than $t' = 0$ (see Fig. 2 in the main text) and $t' = -0.2$, as is the maximum uncertainty ($0.01t$ compared to $0.03t$ and

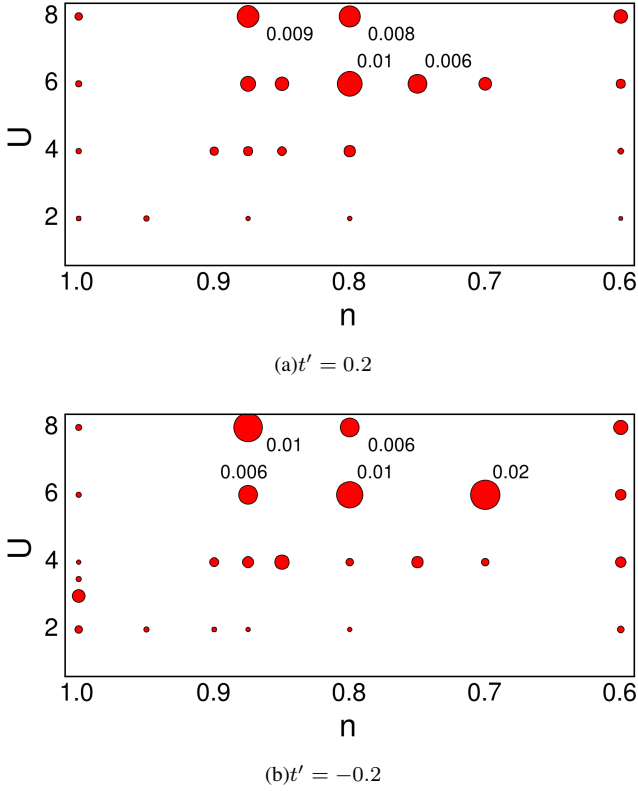

(a) $t' = 0.2$

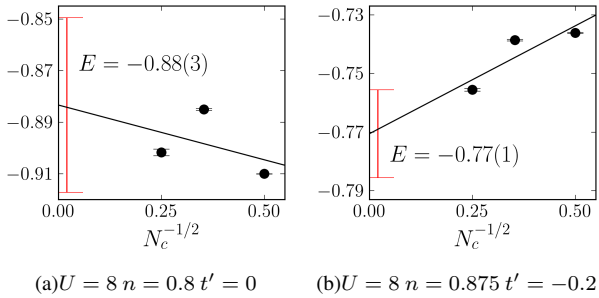

(b) $t' = -0.2$

FIG. 6. DMET energy uncertainty plot for the frustrated Hubbard model with $t' = \pm 0.2$. Refer to Fig. 2 in the main text for the legend.

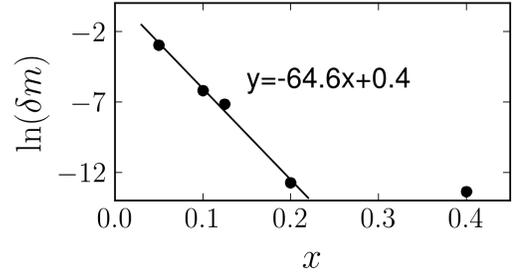

(a) $U = 8$ $n = 0.8$ $t' = 0$     (b) $U = 8$ $n = 0.875$ $t' = -0.2$

FIG. 7. Examples of thermodynamic extrapolations where the energy is sensitive to cluster shape.

$0.02t$, respectively). As mentioned before, the main source of error is the cluster size extrapolation. Two examples of large uncertainties due to cluster size (and shape) effect are shown in Fig. 7. The largest uncertainties are observed at $U = 6$ and moderate doping.

## III. FURTHER RESULTS

In this section, we will expand on the determination of the phase diagram (Fig. 3 in the main text).

The staggered magnetization for the frustrated Hubbard

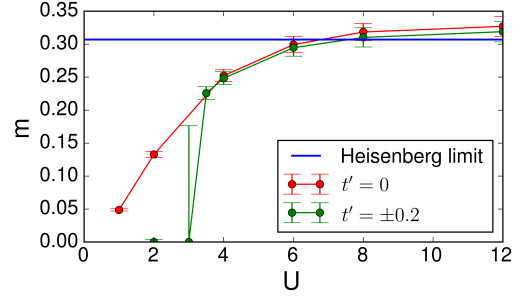

FIG. 8. Staggered magnetization ($m$) of the half-filled Hubbard model for $t' = \pm 0.2$ and $t' = 0$.

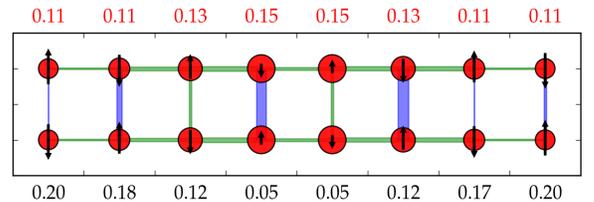

FIG. 9. At $U = 2$, the uncertainty of antiferromagnetic order parameter decreases exponentially with doping. The exponent is $65 \pm 4$.

model at half-filling (compared to the $t' = 0$ model and the Heisenberg limit) is shown in Fig. 8. Due to particle-hole symmetry, the plot is identical for $t' = \pm 0.2$. The onset of antiferromagnetism is at finite $U$ in the frustrated model, between $U = 2$ and 3.5, consistent with previous quantum Monte Carlo simulations[11]. The large error bar at $U = 3$ indicates the sensitivity to impurity cluster sizes near the phase boundary, resulting in a large uncertainty in the thermodynamic extrapolation.

At weak coupling $U = 2$, we find that the antiferromagnetism (in the non-frustrated model) is destroyed already at small doping $x = 0.05$, where the staggered magnetization is $m = 0.00 \pm 0.05$. Although the expectation value is 0, the relatively large uncertainty $\delta m$ reflects that short-range spin fluctuations are still significant, although long-range order does not exist. As we increase doping, $\delta m$ decreases exponentially (Fig. 9). At $U = 2$, we do not find d-wave superconductivity, to within numerical precision.

FIG. 10. Inhomogeneous order from $8 \times 2$ cluster calculations at $U = 4$, $t' = -0.2$ $n = 0.875$.

TABLE I. Energy comparison for different 16-site impurity clusters at $U = 4$ and $t' = -0.2$.

| n | $E_{8\times2}$ | $E_{4\times4}$ |
|---|---|---|
| 0.8 | -1.10483(6) | -1.0507(4) |
| 0.85 | -1.0162(1) | -1.020(2) |
| 0.875 | -0.9966(1) | -0.9989(7) |

We now discuss $U = 4$. We have already shown the order parameters, and the observed thermodynamic extrapolated ground state orders are all homogeneous. However, for $t' = -0.2$, the $8 \times 2$ cluster calculations result in an inhomogeneous state at doping $n = 0.8 - 0.875$, although the energy is significantly higher than that obtained with the $4 \times 4$ clusters at the same fillings. An example of an inhomogeneous pattern is shown in Fig. 10, where one can see a pair density wave and incommensurate magnetic order. In Table I, we compare the energies between the $8 \times 2$ cluster and $4 \times 4$ cluster results at relevant points in the phase diagram for $U = 4$. In all these cases, the $8 \times 2$ cluster has a higher energy, suggesting that the ground state at $U = 4$ is homogeneous or inhomogeneous with a very long wavelength that does not fit in our cluster shapes.

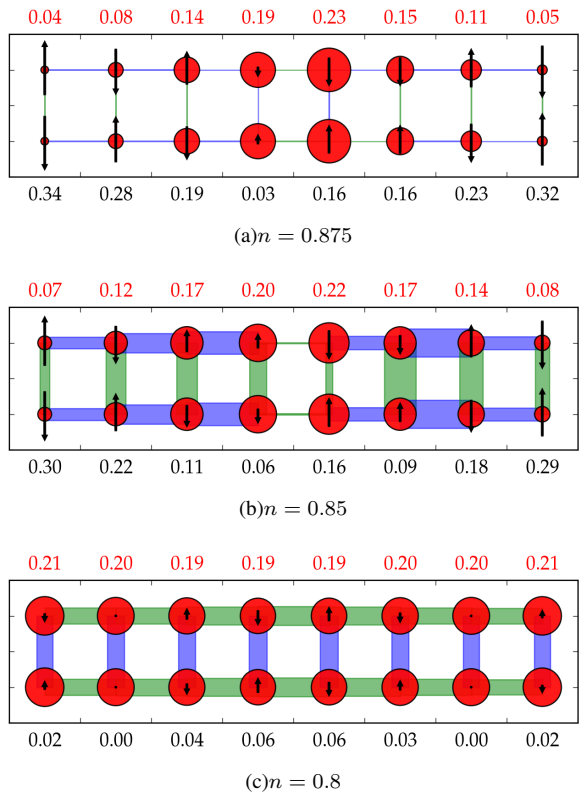

FIG. 11. Inhomogeneous order from $8 \times 2$ cluster calculations at $U = 6$ and $t' = 0$ with fillings 0.875 to 0.8.

At $U = 6$, more interesting inhomogeneous orders start to appear. At $t' = 0$, $8 \times 2$ clusters result in various orders (Fig. 11). At both $n = 0.875$ and $n = 0.85$, $4 \times 4$ clusters are

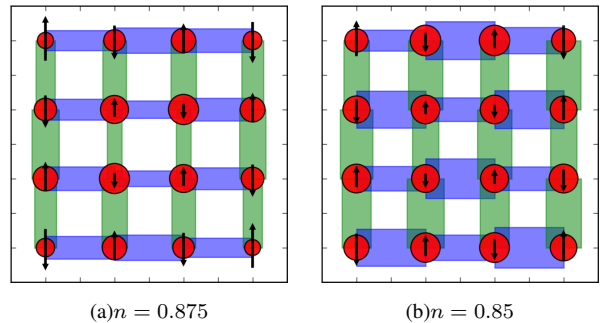

FIG. 12. Local order parameters from $4 \times 4$ cluster calculations at $U = 6$ and $t' = 0$ with fillings 0.875 and 0.85.

significantly lower in energy, suggesting the charge, spin and pairing orders shown in Fig. 11(a) and 11(b) are not stable. At $n = 0.875$, a homogeneous solution with both superconductivity and antiferromagnetism is found (Fig. 12(a)). However, the thermodynamic extrapolation gives zero for both AF and SC order parameters. At $n = 0.85$, the $4 \times 4$ cluster result also shows slight inhomogeneity, with a $(\pi, \pi)$ modulation of the d-wave order parameter (Fig. 12(b)). At $n = 0.8$, where the $8 \times 2$ impurity cluster gives a slightly lower energy ($\Delta E = 0.003(2)$), where DMET calculations indicate a weak spin density wave (Fig. 11(c)). This spin density wave may still exist in the thermodynamic limit because the amplitude is comparable to the staggered magnetization in smaller clusters (eg. $m = 0.04$ for $2 \times 2$ clusters).

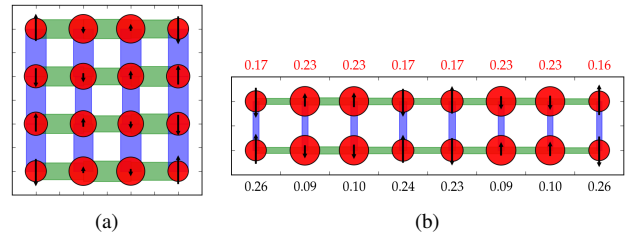

FIG. 13. Local order parameters for $U = 6, n = 0.8, t' = -0.2$.

We now turn to $t' = -0.2$. At $n = 0.8$ and $0.875$, $8 \times 2$ cluster calculations show inhomogeneous orders. At $n = 0.875$, the pattern is similar to what we observed for $t' = 0$ at the same filling, and its energy $E_{8\times2} = -0.8402(4)$ is much higher than that of the $4 \times 4$ homogeneous solution $E_{4\times4} = -0.850(3)$. At $n = 0.8$ (Fig. 13), both $4 \times 4$ and $8 \times 2$ cluster calculations show $\pi$-phase shifts in the spin density and d-wave order, while the $8 \times 2$ cluster has an additional charge density wave. They are very similar in energy, with $E_{8\times2} = -0.9283(2)$ and $E_{4\times4} = -0.927(3)$. This suggests that the ground state here is superconducting with a superimposed spin density wave.

Most results for the underdoped region at $U = 8$ are already shown in the main text (Fig. 5). In Table II, we compare energies for the two 16-site clusters. At all the points shown in the table, the $8 \times 2$ cluster gives a lower energy. An unusual

result is that at $n = 0.8, t' = 0$, the $8 \times 2$ cluster shows a *homogeneous* solution, while both the $4 \times 4$ and $4 \times 2$ clusters give a spin $\pi$-phase shift. This unusual behaviour, where the $8 \times 2$ solution favours homogeneity, is related to the large error in the energy at this point in the thermodynamic extrapolation ($\delta E = 0.03$).

TABLE II. Energy comparison for different 16-site impurity clusters at $U = 8$.

| $t'$ | n | $E_{8 \times 2}$ | $E_{4 \times 4}$ |
|------|-----|------------------|------------------|
| 0 | 0.8 | -0.9018(13) | -0.873(6) |
| 0 | 0.875 | -0.7548(4) | -0.748(4) |
| -0.2 | 0.8 | -0.8487(4) | -0.846(10) |
| -0.2 | 0.875 | -0.7556(5) | -0.737(7)[a] |

[a] The error estimate may not be reliable at this point, because we have only two self-consistent DMET calculations with $M$=1000 and 1200.

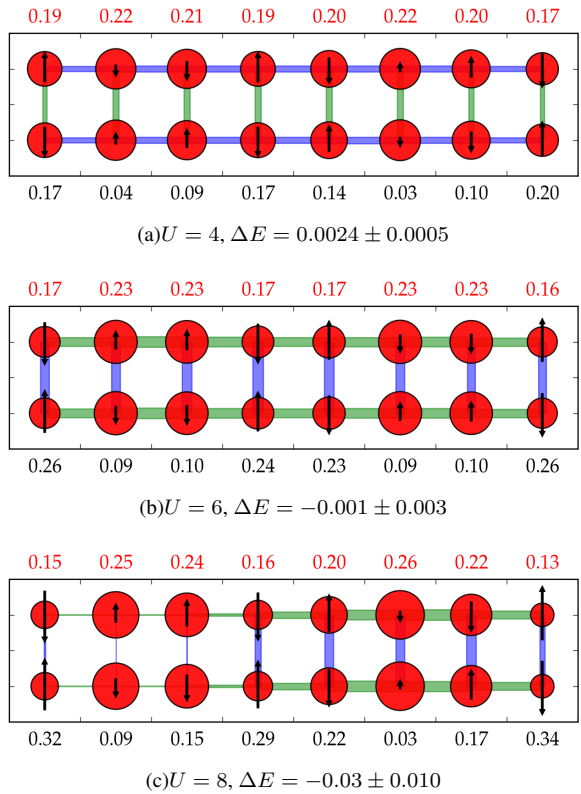

(a) $U = 4$, $\Delta E = 0.0024 \pm 0.0005$

(b) $U = 6$, $\Delta E = -0.001 \pm 0.003$

(c) $U = 8$, $\Delta E = -0.03 \pm 0.010$

FIG. 14. Local order parameters for $n = 0.8, t' = -0.2$ with different $U/t$. $\Delta E = E_{8 \times 2} - E_{4 \times 4}$ is the energy difference between the two 16-site clusters.

In the above we discussed competing orders at different coupling strength, but it is also interesting to look at their evolution with $U$, as shown in Fig. 14. When $U$ increases from 4 to 8, we see increasing charge and spin inhomogeneity although they all show the same pattern of charge localization and a spin $\pi$-phase shift. The d-wave pairing strength first increases and then becomes inhomogeneous. The energy difference between the $8 \times 2$ and $4 \times 4$ clusters also changes

monotonically, although the large error bars prevent us from definitively determining the true order at $U$ =6 and 8.

Finally, we end our discussion on the results by showing the energies across the phase space in Fig. 15. At half-filling, the energy in the frustrated model $t' = \pm 0.2$ is slightly below $t' = 0$, while the difference becomes negligible at large $U$. At the large doping, eg. $n \leq 0.8$, the energy order is dominated by the kinetic effects, i.e. $E_{t'=-0.2} > E_{t'=0} > E_{t'=-0.2}$. The energy curves show more complicate characteristics at underdoped region, especially for $t' = 0$ and $t' = -0.2$.

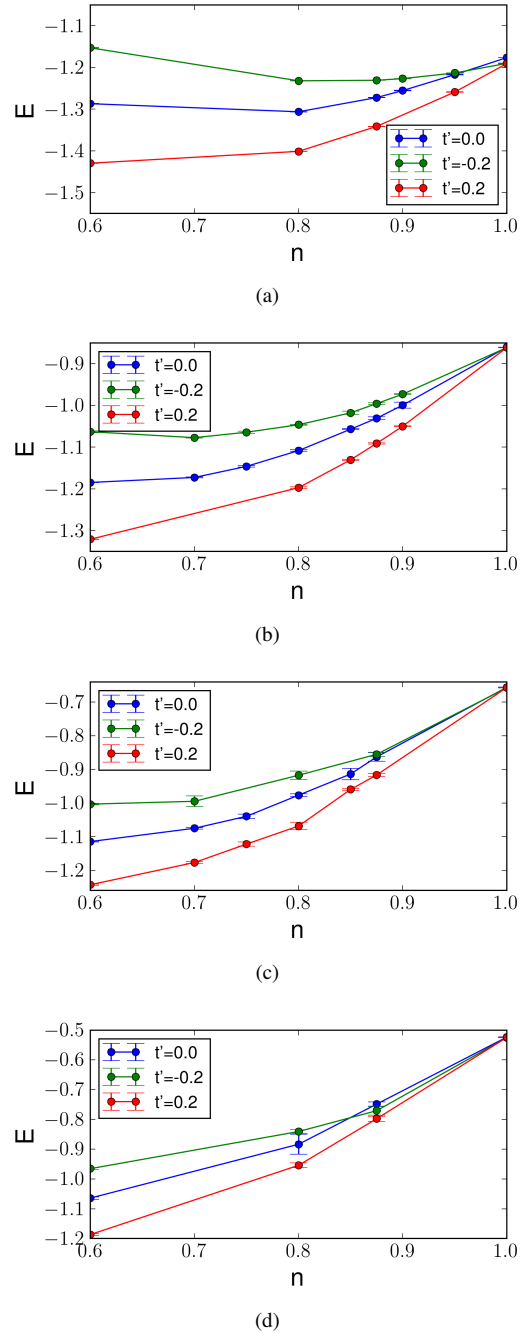

FIG. 15. DMET thermodynamic energy over the phase space. (a) U=2 (b) U=4 (c) U=6 (d) U=8

## IV.  DATA SET

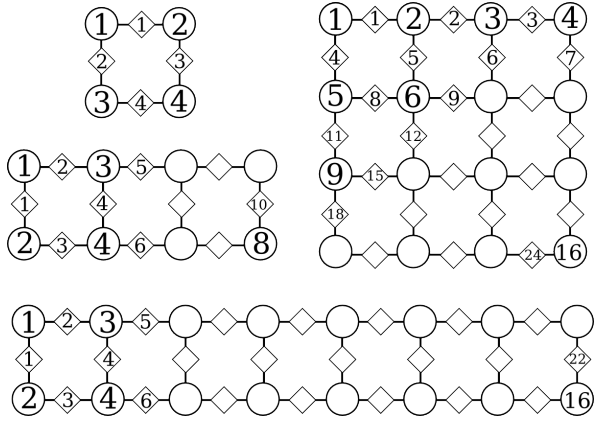

FIG. 16.  The encoding of local order parameters for all impurity clusters. Numbers shown in the circles represent the order of sites, which is associated with charge density and spin density. The numbers in the rhombus represent the order of bonds, or pairs between neighbor sites, which is associated pairing strength. Some numbers are omitted since it is easy to know what they are.

In the attached *TDL.csv* file, we present the energy, chemical potential and (averaged) order parameters computed and their uncertainties at the thermodynamic limit. Since the averaged order parameters are meaningless when inhomogeneity dominates, we have removed these entries from the table.

In the file *clusters.csv*, we present the results for finite impurity clusters. In addition to the results available at thermodynamic limit, we also present the local order parameters. The local order parameters are encoded in an 1D array, which is explained in Fig. The errors shown only include the DMET convergence error, as the other sources of error can be deduced using the procedures described above, from the raw data. We also include the local orders (charge, spin and pairing strength) in this table as a 1D array. The order of the sites and pairs are shown in Fig. 16.

## V.  APPENDIX

Here we prove the equivalence of the Fock spaces spanned by $C_1$ and $C_2$ in the construction of the impurity Schmidt subspace, as defined in section (I C). Precisely, we need to prove

1. $C_2$ is orthonormal, $C_2^T C_2 = I$. (It is easy to see $C_1$ is orthonormal, because $Q$ is a unitary matrix from QR decomposition.)

2. $C_2 = C_1 V$, which is equivalent to $C_1^T C_2 = V$, where $V$ is unitary.

To prove (1) $C_2^T C_2 = I$, we need the idempotency of density matrices $\rho^2 = \rho$. Considering only the upper-left block of $\rho$, we have

$$\rho_{\text{imp}}^2 + \rho_c^T \rho_c = \rho_{\text{imp}} \tag{33}$$

From Eq. (13) and (14), we know $A' = U(I - \Lambda)^{\frac{1}{2}}, B' = -\rho_c(A'^T)^{-1}$. Therefore

$$C_2^T C_2 = \begin{pmatrix} A^T & A^{-1}\rho_c^T \\ A'^T & -(A')^{-1}\rho_c^T \end{pmatrix} \begin{pmatrix} A & A' \\ \rho_c(A^T)^{-1} & -\rho_c(A'^T)^{-1} \end{pmatrix} = \begin{pmatrix} A^T A + A^{-1}\rho_c^T \rho_c(A^T)^{-1} & A^T A' - A^{-1}\rho_c^T \rho_c(A'^T)^{-1} \\ A'^T A - (A')^{-1}\rho_c^T \rho_c(A^T)^{-1} & A'^T A' + (A')^{-1}\rho_c^T \rho_c(A'^T)^{-1} \end{pmatrix}$$
$$= \begin{pmatrix} \Lambda + \Lambda^{-\frac{1}{2}}\Lambda(I - \Lambda)\Lambda^{-\frac{1}{2}} & \Lambda^{\frac{1}{2}}(I - \Lambda)^{\frac{1}{2}} - \Lambda^{-\frac{1}{2}}\Lambda(I - \Lambda)(I - \Lambda)^{-\frac{1}{2}} \\ (I - \Lambda)^{\frac{1}{2}}\Lambda^{\frac{1}{2}} - (I - \Lambda)^{-\frac{1}{2}}\Lambda(I - \Lambda)\Lambda^{-\frac{1}{2}} & I - \Lambda + (I - \Lambda)^{-\frac{1}{2}}\Lambda(I - \Lambda)(I - \Lambda)^{-\frac{1}{2}} \end{pmatrix} = I \tag{34}$$

For (2), since

$$V = C_1^T C_2 = \begin{pmatrix} A & A' \\ Q^T \rho_c(A^T)^{-1} & -Q^T \rho_c(A'^T)^{-1} \end{pmatrix} \tag{35}$$

we have

$$VV^T = \begin{pmatrix} A & A' \\ Q^T \rho_c(A^T)^{-1} & -Q^T \rho_c(A'^T)^{-1} \end{pmatrix} \begin{pmatrix} A^T & A^{-1}\rho_c^T Q \\ A'^T & -A'^{-1}\rho_c^T Q \end{pmatrix}$$
$$= \begin{pmatrix} AA^T + A'A'^T & \rho_c^T Q - \rho_c^T Q \\ Q^T \rho_c - Q^T \rho_c & Q^T \rho_c(AA^T)^{-1}\rho_c^T Q + Q^T \rho_c(A'A'^T)^{-1}\rho_c^T Q \end{pmatrix} = \begin{pmatrix} I & 0 \\ 0 & R[A^{-1} + (I - A)^{-1}]R^T \end{pmatrix} \tag{36}$$

In the bottom-right block

$$A^{-1} + (I - A)^{-1} = U\Lambda^{-1}U^T + U(I - \Lambda)^{-1}U^T$$
$$= U\Lambda^{-1}(I - \Lambda)^{-1}U^T$$
$$= [A(I - A)]^{-1}$$
$$= (B^T B)^{-1} = (R^T R)^{-1} = R^{-1}(R^T)^{-1} \tag{37}$$

So $VV^T = I$. Here we assume $R$ is invertible, which is true if and only if we have the full set of $N_c$ bath orbitals coupled to the impurity. This is generally true in lattice settings where the impurity and the environment are strongly coupled. Some-

times the bath can be smaller than the impurity in molecules and when we use a large basis set, and in these cases, special treatment is needed.

[1] G. Knizia and G. K.-L. Chan, Phys. Rev. Lett. **109**, 186404 (2012).

[2] P.-G. de Gennes, *Superconductivity Of Metals And Alloys* (Benjamin, New York, 1966).

[3] D. Yamaki, T. Ohsaku, H. Nagao, and K. Yamaguchi, International Journal of Quantum Chemistry **96**, 10 (2004).

[4] J. Bardeen, L. N. Cooper, and J. R. Schrieffer, Phys. Rev. **108**, 1175 (1957).

[5] G. Knizia and G. K.-L. Chan, Journal of Chemical Theory and Computation **9**, 1428 (2013).

[6] S. Datta and P. F. Bagwell, Superlattices and Microstructures **25**, 1233 (1999).

[7] G. K.-L. Chan and M. Head-Gordon, The Journal of chemical physics **116**, 4462 (2002).

[8] G. K.-L. Chan, The Journal of chemical physics **120**, 3172 (2004).

[9] S. Sharma and G. K.-L. Chan, The Journal of chemical physics **136**, 124121 (2012).

[10] I. W. Bulik, G. E. Scuseria, and J. Dukelsky, Phys. Rev. B **89**, 035140 (2014).

[11] H. Q. Lin and J. E. Hirsch, Phys. Rev. B **35**, 3359 (1987).


\end{widetext}
\end{document}